\documentclass[]{imsart}

\RequirePackage{amsthm,amsmath,amsfonts,amssymb}
\RequirePackage{natbib}

\usepackage{amsmath}
\usepackage{amssymb}
\usepackage{amsthm}
\usepackage{bm}

\usepackage[table,xcdraw]{xcolor}
\usepackage{xcolor}

\usepackage{tikz-cd}
\usepackage{tikz}
\usetikzlibrary{shapes.geometric, arrows}
\tikzstyle{startstop} = [rectangle, rounded corners, minimum width=3cm, minimum height=1cm,text centered, draw=black, fill=red!10]
\tikzstyle{arrow} = [thick,->,>=stealth]

\usetikzlibrary{bayesnet}

\usepackage{hyperref}

\usepackage{graphicx}
\graphicspath{ {./images/} }

\newcommand{\blambda}{{\bm{\lambda}}}
\newcommand{\btheta}{{\bm{\theta}}}
\newcommand{\bpsi}{{\bm{\psi}}}
\newcommand{\by}{\mathbf{y}}

\begin{document}

\begin{frontmatter}
\title{Athlete rating in multi-competitor games with scored outcomes via monotone transformations}

\begin{aug}
\author{\fnms{Jonathan}~\snm{Che}}
\and
\author{\fnms{Mark}~\snm{Glickman}}
\address{Harvard University}
\end{aug}

\begin{abstract}

Sports organizations often want to estimate athlete strengths.
For games with scored outcomes, a common approach is to
assume observed game scores follow a normal distribution conditional
on athletes' latent abilities, which may change over time.
In many games, however, this assumption of conditional 
normality does not hold.
To estimate athletes' time-varying latent abilities using non-normal game score data, we propose a Bayesian dynamic linear model with flexible monotone response transformations.
Our model learns nonlinear monotone transformations to address non-normality in athlete scores and can be easily fit using standard regression and optimization routines, which we implement in the \texttt{dlmt} package in \texttt{R}.
We demonstrate our method on data from several Olympic sports, including biathlon, diving, rugby, and fencing.

\end{abstract}


\end{frontmatter}

\section{Introduction}

Sports and gaming organizations, athletes, and fans often wish to estimate how good athletes are at their sports.
This problem of athlete rating can affect planning and preparation for games, at both organizational and individual levels.
For example, ratings may influence how sports organizations allocate resources across their athletes prior to an event to maximize their chances of winning.
Ratings may also be used for designing tournaments and league play.
This includes pairing athletes with similar estimated abilities in head-to-head competitions, dividing a large set of competitors into smaller tournaments according to skill level, or selecting top athletes to compete in elite ``by invitation only'' events.

Approaches that rely on 
statistical modeling enable researchers to infer athletes' abilities from their game outcomes in a principled manner.
A typical method for constructing athlete ratings treats each athlete's strength as a latent parameter (or vector of parameters) within a probability model.
The models then use observed game outcomes to estimate the latent ability parameters, which may be thought of as fixed or varying over time.

A variety of methods for estimating dynamic (i.e., time-varying) athlete ratings have been proposed for games with binary (i.e., win/loss) and rank-ordered outcomes.
For multi-competitor games, approaches to rating competitors typically extend the Plackett-Luce model \citep{plackett1975analysis} through the evolution of the latent ability parameters.
For example, \citet{glickman2015stochastic} models the evolution as a discrete stochastic process, \citet{caron2012bayesian} uses a nonparametric stochastic process, \citet{baker2015golf} interpolates abilities between discrete time points, and \citet{mckeough2020tale} considers parametric growth curves over time.
Dynamic models have also been proposed for head-to-head games with win/loss outcomes, in both team \citep{herbrich2006trueskill} and individual \citep{glickman1999parameter, glickman2001dynamic, cattelan2013dynamic, baker2015time} game settings.

Score outcomes are more granular than rank-order outcomes; as a result, models that effectively use score data may outperform models that only use ranking data.
Observed rank outcomes can be viewed as partially censored score outcomes.
For example, in a race, each runner's performance can be recorded as a race time, which can then be mapped into a race placement.
In this instance, the race placement is the ranking and the race time is the score.
Score data can potentially provide information, particularly about the gaps in performances between athletes, which may be useful to a rating model.

While various dynamic models that use score or score-related information have been proposed for head-to-head games \citep{harville1977use, glickman1998state, lopez2018often, ingram2019point, kovalchik2020extension}, we are unaware of similar work for multi-competitor games.
In this paper, we extend the normal dynamic linear model (DLM) proposed by \citet{harville1977use} and \citet{glickman1998state}, who 
model NFL football game scores using DLMs, to rate athletes who compete in multi-competitor games.

DLMs have a rich history of use in fields spanning engineering, finance, ecology, and medicine, among many others (e.g., see \citet{auger2021introduction, zhou2020semiparametric, wang2019modeling, hotz2018predicting} for some recent examples).
They provide a simple, natural framework for 
modeling time-varying outcomes through parameters that are assumed
to follow a latent stochastic process.
In the context of sports, they assume that an athlete's latent ability varies across time periods as a discrete stochastic process, and that an athlete's scores are normally distributed around their latent ability parameter within each time period.

Athletes' game scores, however, can often be heavy-tailed or skewed, which suggests that the normal assumption in DLMs may not hold.
Furthermore, games with blowouts or close wins may produce scores that do not accurately reflect athletes' skills.
To directly account for blowouts, \citet{harville2003selection} considers simple strategies such as capping margins of victory and adjusting extreme scores using hazard functions.

In this paper, we improve on this idea by directly learning a flexible transformation of the scores from the data.
We propose a Bayesian DLM with a monotone spline outcome transformation \citep{Ramsay1988} to rate athletes who compete in multi-competitor games with scored outcomes.
Our model uses a Bayesian approach to learn the best transformation from the data in a principled manner.
Once the transformation has been estimated, we apply a DLM to the transformed data, thus addressing the non-normality of the score outcomes while still preserving the efficiency and transparency of standard methods for time-series data.
We implement our model in the \texttt{dlmt} package in \texttt{R}.\footnote{Available at \href{https://github.com/jche/dlmt}{https://github.com/jche/dlmt}.}

DLMs are classically used in applications where fixed, known (or assumed) processes dictate both how latent parameters evolve over time and how observations are generated from latent parameters.
For example, DLMs have been applied to trajectory estimation, where physics equations determine how an object's true position varies over time.
In these settings, much work has been done with Extended and Unscented Kalman Filters \citep{wan2000unscented}, which allow for nonlinearities in these relationships.

In our setting, however, it is unclear a priori how to fix an assumed nonlinear relationship between athletes' latent abilities and their game scores;
as such, this relationship must be learned from the data.
To learn this relationship, early research on DLMs considered learning an optimal Box-Cox transformation of the data, either by using a naive grid search \citep{lenk1990transformations} or by using score-based methods \citep{atkinson1996deletion}.
While these methods are straightforward and intuitive, the Box-Cox transformation lacks the flexibility required for many real-world settings.
Recent work has considered using more flexible spline transformations, albeit in different settings from the one in this paper;
\citet{Xia2000a} uses a monotone spline transformation in the simpler setting of univariate autoregression with static parameters for an ecological time-series dataset, and \citet{king2021warped} utilizes a spline-based transformation for the analysis of count data in a DLM setting.
We extend this line of work by applying a flexible monotone spline transformation to the case of multicompetitor game scores in a DLM setting, where athletes may compete in any number of games within each time period.

The paper proceeds as follows.
Section \ref{sec:model} describes the general DLM framework and demonstrates the incorporation of monotone transformations into the model.
Section \ref{sec:fitting} describes an efficient model-fitting algorithm for the DLM with transformations.
Finally, Section \ref{sec:results} compares our DLM to other candidate models for rating athletes in multi-competitor games, using data provided by the US Olympic and Paralympic Committee.

\section{Model}
\label{sec:model}

We introduce the model proposed in this paper in two stages.
First, we present a standard DLM framework for rating athletes.
We then modify it to account for game effects and non-normal outcomes, and address the special case of head-to-head games.

\subsection{Standard DLM for athlete rating}

A standard DLM for rating athletes models each athlete's observed scores as normally distributed around their latent ability, which evolves over time.
The model likelihood for the score observed from a single athlete competing during time $t$ takes the form:
\begin{equation}
    p(y_t \mid \theta_t, \sigma^2) = N(y_t \mid \theta_t, \sigma^2),
    \label{eq:lik1}
\end{equation}
where $y_t$ is the observed score, $\theta_t$ is the athlete's latent ability parameter at time $t$, and $\sigma^2$ is the observation variance.
In this paper, we allow the latent ability parameter to evolve between time points as a normal random walk:
\begin{equation}
    p(\theta_{t+1} \mid \theta_t, \sigma^2, w) = N(\theta_{t+1} \mid \theta_t, \sigma^2w),
    \label{eq:innov1}
\end{equation}
where $w$ is an additional parameter that controls how much latent abilities may vary over time relative to the observation variance.
Other stochastic processes may also be considered for the evolution of latent ability parameters, such as a mean-preserving random walk \citep{glickman2015stochastic} or an autoregressive process \citep{glickman1998state}.
In this paper, we use a simple normal random walk rather than an autoregressive process to acknowledge that weak athletes' latent abilities are not likely to improve, i.e., revert to the mean, when they avoid playing games over several time periods.
Unlike these other stochastic processes, the normal random walk diverges over time; 
this increasing variance represents the increasing uncertainty about each athlete's latent ability if they do not play any games.
In practice, we place an upper limit on the variance of athlete latent abilities to prevent them from growing too large (see Appendix \ref{AppA_imp} for details).

In a general sporting setting, we divide time into discrete \textit{rating periods}, with multiple games within each rating period and multiple athletes within each game.
The choice of rating period should reflect the structure of the data.
Our model assumes that athletes' latent abilities remain constant within each rating period, but that they may vary between rating periods. 
Sports typically have regularly spaced seasons which serve as natural rating period breakpoints; for example, the biathlon season spans November through March, so we may assign the full season to a single rating period or split each season into two rating periods to allow mid-season changes in ability.
Shorter rating periods generally reduce prediction bias, since more recent data is used for prediction, but they increase variance, since fewer games are used to infer latent abilities within each period.
Appendix \ref{AppA_ratingperiodsens} provides further guidance on this bias-variance tradeoff and illustrates how different choices of rating periods marginally affect results on real data.

Let $p$ denote the total number of athletes in the data.
Let $T$ denote the total number of discrete rating periods, indexed by $t = 1, \dots, T$.
Finally, let $n_t$ denote the total number of observed scores within rating period $t$.
Athletes may compete in any number (including zero) of games within each rating period.
The data model (Equation \ref{eq:lik1}) and innovation model (Equation \ref{eq:innov1}) may now be rewritten in multivariate form as:
\begin{align}
	p(\by_{t} \mid \btheta_t, \sigma^2) 
	    &= N(\by_t \mid X_{t} \btheta_t, \sigma^2 I_{n_t}) \nonumber \\ 
	p(\btheta_{t+1} \mid \btheta_{t}, \sigma^2, w) 
	    &= N(\btheta_{t+1} \mid \btheta_{t}, \sigma^2 w I_p). \nonumber 
\end{align}
Here, $\by_t$ is the $n_t \times 1$ column vector of observed scores, $\btheta_t$ is the $p \times 1$ column vector of athlete latent abilities for rating period $t$, and $I_k$ denotes the $k \times k$ identity matrix.
The $n_t \times p$ model matrix $X_t$ simply matches each athlete's observed score(s) to their latent ability parameter.\footnote{Temporarily suppressing the time subscript $t$, the matrix $X$ has a single nonzero entry per row.
Per row $r$, if entry $r$ of the column vector $\by$ corresponds to a score earned by athlete $a$, then $X_{ra}$ is set to equal 1.}

\subsection{Addressing game effects}


In practice, athlete-rating DLMs need to account for conditions that affect game scores in a manner unrelated to latent athlete abilities.
For example, a hot day might make all athletes in a race run more slowly, but the increased race times do not indicate weaker athletes.
One approach to incorporate game-specific variation is to assume game-specific intercepts as part of the outcome model.
This approach has been used, for example, by \cite{glickman1998state}.

Instead of assuming game-specific intercepts, we pre-process the data by subtracting game-specific means from the observed scores.
This approach avoids concerns relating to the arbitrary specification of priors for the intercepts, which could affect downstream results.
Each game-centered score $\tilde{y}_{tg}$ for an individual athlete in game $g$ within rating period $t$ may then be modeled 
using a game-centered latent ability, as:
\begin{equation}
	p(\tilde{y}_{tg} \mid \theta_{t}, \sigma^2)
	= N(\tilde{y}_{tg} \mid \theta_t - \bar{\theta}_{tg}, \sigma^2). \nonumber 
\end{equation}
The value $\bar{\theta}_{tg}$ denotes the average latent ability across all of the players in game $g$ within rating period $t$.
Subtracting $\bar{\theta}_{tg}$ from each athlete's latent ability adjusts for the fact that competing against a stronger pool of opponents in a game naturally results in worse scores relative to the competition.

The variation in scores may also differ across games within a sport.
For example, biathlon races vary in length.
Longer races give strong athletes more time to build a larger lead and weak athletes more time to fall behind, increasing the observed variation in scores.
Ideally, the transformation introduced in the following section would address this issue by shrinking extreme game-centered scores toward the mean.
If the variation in scores significantly differs from game to game in a particular application, however, it may be reasonable to additionally pre-process the scores 
e.g., by log-transforming them, prior to game-centering, or (for large multicompetitor sports) by game-centering and then scaling each game's scores to have variance 1.

To simplify notation for vector of pre-processed, game-centered scores $\tilde{\by}_t$, we write:
\begin{equation}
	p(\tilde{\by}_{t} \mid \btheta_{t}, \sigma^2) = N(\tilde{\by}_{t} \mid \bar{X}_t \btheta_{t}, \sigma^2 I_{n_t}), \label{eq:lik4}
\end{equation}
where 
$ \tilde{\by}_t \equiv
\begin{bmatrix}
    \tilde{\by}_{t1} \\
    \vdots \\
    \tilde{\by}_{tq}
\end{bmatrix}
$
and
$\bar{X}_t \equiv 
\begin{bmatrix}
    H_{n_{t1}} X_{t1} \\
    \vdots \\
    H_{n_{tg_t}} X_{tg_t}
\end{bmatrix}$,
for centering matrices $H_k \equiv I_{k} - \mathbf{1}_{k} \mathbf{1}_{k}^T$ with
$\mathbf{1}_{k}$ denoting the $k \times 1$ column vector with entries equal to one.
For games $g = 1, \dots, g_t$ within rating period $t$, the $n_g \times 1$ vector of scores and $n_g \times p$ model matrix for game $g$ are written as $\tilde{\by}_{tg}$ and $X_{tg}$, respectively.


\subsection{Addressing non-normal outcomes}

In many games, we might also suspect that athletes' game-centered scores are not normally distributed around their game-centered latent abilities as assumed by Equation \ref{eq:lik4}.
Instead, we may assume that some transformation of the athletes' scores is normally distributed around their latent abilities, so that the resulting model for transformed outcomes is:
\begin{equation}
	p(\tau_\blambda(\tilde{\by}_t) \mid \btheta_{t}, \sigma^2, \blambda) 
    = N(\tau_\blambda(\tilde{\by}_t) \mid \bar{X}_t \btheta_{t}, \sigma^2 I_{n_t}). \nonumber
\end{equation}
The transformation $\tau_\blambda(\cdot)$ is a function parameterized by a vector-valued parameter $\blambda$.
While the above assumption is likely reasonable for most sports, we note that it may be difficult to justify for sports with very low scores, e.g., soccer or hockey.

In this paper, we use the monotone spline transformation from \cite{Ramsay1988}, but any monotone transformation with a computable Jacobian would work as well.
The monotone spline transformation is a polynomial spline built from an I-spline basis (see \cite{Ramsay1988} for more detail).
For a given polynomial order $d$ and knot sequence $\mathbf{k}$, an I-spline basis consists of $B$ fixed, monotonically increasing basis functions $I_b(y \mid d, \mathbf{k})$, $b=1,\ldots,B$.
The monotone spline transformation is then constructed as a linear combination of these basis functions:
\begin{equation}
	\tau_{\blambda}^{MS}(y) = \lambda_0 + \sum_{b=1}^B \lambda_b I_b(y \mid d, \mathbf{k}), \nonumber
\end{equation}
where $\lambda_0$ represents an intercept term and $\lambda_1, \dots, \lambda_B$ determine the shape of the transformation.
We treat $\lambda_0$ as fixed, and define the transformation parameter $\blambda$ as
$\blambda \equiv 
    \begin{bmatrix}
        \lambda_1 \ \dots \ \lambda_B
    \end{bmatrix}^T$.
Note that because each basis function $I_b(\cdot)$ is monotone increasing, constraining the parameters $\lambda_1, \dots, \lambda_B$ to be non-negative ensures that the resulting spline transformation is also monotone increasing.
Also, the sum $\sum_{b=1}^B \lambda_b$ determines the range of the monotone spline transformation function, so we constrain it to equal a constant $c$ for identifiability.

Choosing the number and placement of the spline knots is generally a challenging problem.
In this paper, we simply place three interior knots at evenly spaced quantiles \citep[Section 7.4.4]{james2013introduction}.
Adding more knots increases the flexibility of the transformation, but also increases the number of parameters that will need to be optimized in our model-fitting procedure.
In our application, we found that three interior knots provided sufficient flexibility while maintaining computational efficiency and stability.
Appendix \ref{AppA_imp} provides further guidance about knot choice.

\subsection{Full Bayesian model}

The full DLM with transformations is therefore as follows:
\begin{align}
	p(\bpsi_t \mid \btheta_{t}, \sigma^2, \blambda)
	&= N(\bpsi_t \mid \bar{X}_t \btheta_{t}, \sigma^2 I_{n_t})  \label{eq:likfin} \\
	p(\btheta_{t+1} \mid \btheta_{t}, \sigma^2, w)
	&= N(\btheta_{t+1} \mid \btheta_{t}, \sigma^2 w I_p), \label{eq:innovfin}
\end{align}
where we define $\bpsi_t \equiv \tau_\blambda(\tilde{\by}_t)$, the transformed, game-centered observations, suppressing dependence on $\blambda$ to simplify notation.
To complete the model specification, we specify prior distributions for $\btheta_1$, $\sigma^2$, $w$, and $\blambda$:
\begin{align}
    p(\btheta_1 \mid \sigma^2, v_0)
    	&= N(\btheta_1 \mid 0, \sigma^2 v_0 I_p) \nonumber \\
    p(\sigma^2)
    	&= \text{Inv-Gamma}(\sigma^2 \mid a_0, b_0) \nonumber \\
	p(w)
	    &= \text{Half-Normal}(w \mid s_w^2)  \nonumber\\ 
	p(\blambda)
	   &= c \cdot \text{Dirichlet}(\blambda \mid \bm{\alpha}).  \nonumber
\end{align}
The hyperparameters $v_0$, $a_0$, $b_0$, $s_w$, $c$, and $\bm{\alpha}$ may be set to reflect prior beliefs about $\btheta_1$, $\sigma^2$, $w$, and $\blambda$.
In the absence of available prior information, we set $v_0 = 10$, $a_0 = b_0 = 0.1$, and $s_w = 1$ to keep the priors fairly uninformative about $\btheta_1$, $\sigma^2$, and $w$ \citep{gelman1995bayesian}.
We set $\bm{\alpha}$ to be proportional to the $\lambda$ parameters corresponding to the identity transformation, but keep $\sum_{b=1}^B \alpha_b = 1$ to keep the prior diffuse.
If more shrinkage toward the identity transformation is desired, $\sum_{b=1}^B \alpha_b$ can be set to a higher value.
Finally, we set $\lambda_0$ to equal the lowest score in the data and $c$ to equal the range of the scores in the data so that the learned transformation roughly preserves the scale of the original data.

The Dirichlet prior on $\blambda$ constrains the transformation parameter $\blambda$ to have nonnegative components that sum to $c$.
In practice, we find that using a weakly regularizing unconstrained prior for $\blambda$ \citep{kowal2020simultaneous} often results in better performance:
\begin{equation}
    p(\lambda_b) = N^{+}(\lambda_b \mid \alpha_b, s^2_\blambda) \text{ for } b = 1, \dots, B, \nonumber
\end{equation}
where $N^+$ indicates a normal distribution truncated below at 0.
Theoretically, the shape of the monotone spline transformation is unidentifiable without a constraint on $\sum_{b=1}^B \lambda_b$; empirically, the mild regularization induced by the truncated normal priors effectively addresses this issue.
For this unconstrained model, we set $\bm{\alpha}$ equal to the $\lambda$ parameters corresponding to the identity transformation and control shrinkage toward the identity transformation using $s^2_\blambda$.
To allow maximum flexibility, we set $s^2_\blambda$ to a large value.

\begin{figure}[t]
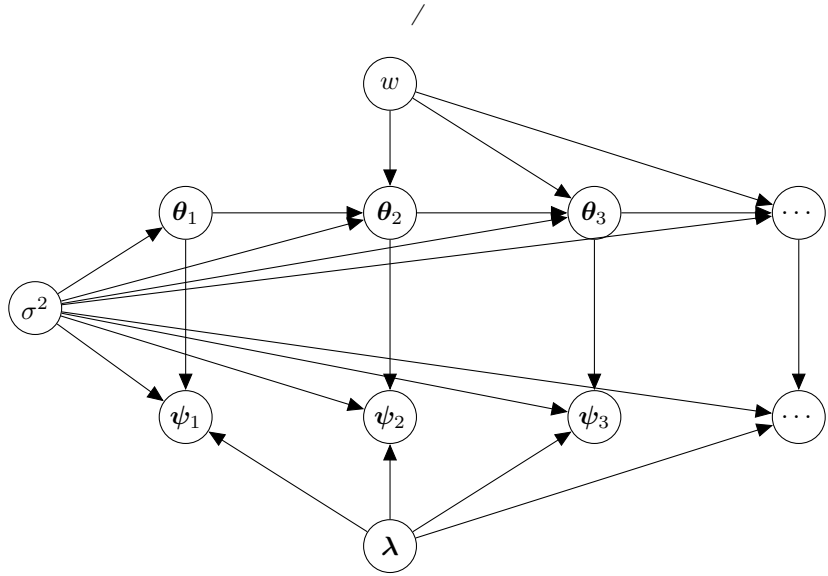

    \centering
    \tikz{ %
    \node[latent] (t1) {$\btheta_1$} ; %
    \node[latent, right=2cm of t1] (t2) {$\btheta_2$} ; %
    \node[latent, right=2cm of t2] (t3) {$\btheta_3$} ; %
    \node[latent, right=2cm of t3] (dots) {$\dots$} ; %
    \node[latent, below=2cm of t1] (p1) {$\bpsi_1$} ; %
    \node[latent, right=2cm of p1] (p2) {$\bpsi_2$} ; %
    \node[latent, right=2cm of p2] (p3) {$\bpsi_3$} ; %
    \node[latent, right=2cm of p3] (dots2) {$\dots$} ; %
    \node[latent, above=of t2] (w) {$w$} ; %
    \node[latent, below=of p2] (lam) {$\blambda$} ; %
    \node[latent, below=0.55cm of t1, xshift=-2cm] (sig) {$\sigma^2$} ; %
    \edge {sig} {t1} ; %
    \edge {sig,w,t1} {t2} ; %
    \edge {sig,w,t2} {t3} ; %
    \edge {sig,w,t3} {dots} ; %
    \edge {sig,lam,t1} {p1} ; %
    \edge {sig,lam,t2} {p2} ; %
    \edge {sig,lam,t3} {p3} ; %
    \edge {sig,lam,dots} {dots2} ; %
  }
    \caption{Graphical model for DLM with transformations}
    \label{fig:graph}
\end{figure}

\begin{figure}[t]
    \centering
    \includegraphics[width=5.5in]{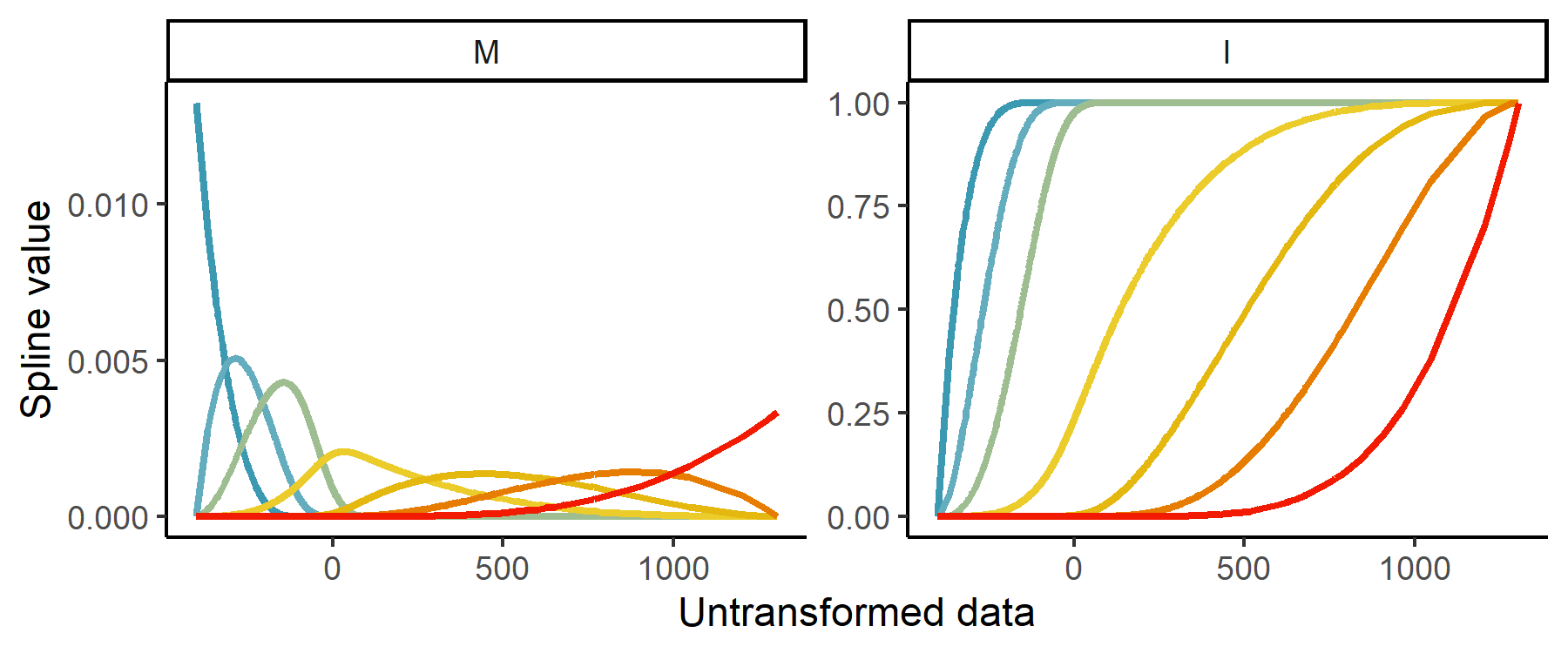}
    \caption{Spline basis functions for biathlon relay training set}
    \label{fig:splines}
\end{figure}

Figure \ref{fig:graph} displays the graphical model representing the relationships between the model parameters $\{\btheta_t\}_{t = 1, \dots, T}$, $\sigma^2$, $w$, and $\blambda$ and the transformed data $\{\bpsi_t\}_{t=1, \dots, T}$.
Figure \ref{fig:graph} displays the conditional independences implied by Equations \ref{eq:likfin} and \ref{eq:innovfin}, which allow us to factor the joint distribution of our untransformed data and model parameters as:
\begin{align}
	&p(\by_{1:T}, \btheta_{1:T}, \sigma^2, w, \blambda) = \nonumber \\
	&J(\bpsi_{1:T} \to \by_{1:T}) \times p(\sigma^2) p(w) p(\blambda)
	\times \prod_{t=1}^{T} p(\bpsi_t \mid \btheta_t, \sigma^2, \blambda)
	\times \prod_{t=1}^{T} p(\btheta_t \mid \btheta_{t-1}, \sigma^2, w), \nonumber
\end{align}
where $J(\bpsi_{1:T} \to \by_{1:T})$ is the Jacobian of the inverse transformation, $\tau_\blambda^{-1}(\cdot)$.
The Jacobian term is vital to appropriately account for how the transformation rescales the data.
For example, the Jacobian of the inverse monotone spline transformation is:
\begin{equation}
	J^{MS}(\psi \to y) = \sum_{b=1}^B \blambda_b M_b(y \mid d, \mathbf{k}). \nonumber
\end{equation}
The I-spline basis functions used for the monotone spline are constructed by integrating M-spline basis functions; here, $M_b(\cdot)$ are the M-spline basis functions corresponding to their respective I-splines.
Figure \ref{fig:splines} displays the seven spline basis functions constructed for the biathlon relay training dataset we study in Section \ref{sec:results}.
Figure \ref{fig:splines} shows how each I-spline basis function is constructed as the integral of an M-spline basis function.

\subsection{Head-to-head games}

While the general multi-competitor setup introduced above can be used for games with any number of players, we introduce a slightly simpler setup for the special case of head-to-head games.
In head-to-head games, it is generally more natural to consider monotone transformations of score differences, which have no direct analogues in the multi-competitor setting.
Instead of modeling the $2g_t \times 1$ vector of athlete scores $\by_t$, we can model the $g_t \times 1$ vector of score differences $\mathbf{z}_t$, where we subtract the second athlete's score from the first athlete's score.
The resulting model likelihood is:
\begin{equation}
    p(\tau_\blambda(\mathbf{z}_t) \mid \btheta_{t}, \sigma^2, \blambda) = N( \tau_\blambda(\mathbf{z}_t) \mid Z_t \btheta_{t}, \sigma^2 I_{n_t}). \nonumber
\end{equation}
Note that we have simply replaced the model matrix $\bar{X}_t$ with the model matrix $Z_t$, which we define as a matrix with two nonzero entries per row: 1 in the column corresponding to the first athlete and -1 in the column corresponding to the second athlete.
Each observed score difference is taken as the first athlete's score minus the second athlete's score, so that each transformed score difference is normally distributed around the latent ability parameter difference between the two competing athletes.

\section{Model fitting}
\label{sec:fitting}

\subsection{Model-fitting procedure}

We estimate the model parameters $(\btheta_{1:T}, \sigma^2, w, \blambda)$ via a two-step procedure.
In our application of interest, we would like to be able to quickly update athlete ratings (i.e., latent ability estimates) shortly after the results from a game.
While we could estimate the full posterior distribution of all of the model parameters after each game, this may be an unnecessarily slow and computationally expensive process.
Instead, we fit the model using the following procedure:
\begin{enumerate}
    \item Estimate $w$ and $\blambda$: using a training subset consisting of the first $T_{train}$ rating periods in the dataset, obtain estimates $\hat{w}$ and $\hat{\blambda}$.
    \item Estimate $\btheta_{1:T}$ and $\sigma^2$: given $\hat{w}$ and $\hat{\blambda}$, use the full dataset to obtain estimates $\hat{\btheta}_{1:T}$ and $\hat{\sigma}^2$.
\end{enumerate}
While step 1 is computationally expensive, step 2 can be implemented efficiently.
When the results from a new game become available, we can just run step 2 using the previously learned values of $\hat{w}$ and $\hat{\blambda}$ to quickly update ratings.

We first describe step 2, the final estimation of $\btheta_{1:T}$ and $\sigma^2$.
Given fixed values for $w$ and $\blambda$, we can transform the full dataset using $t_{\blambda}(\cdot)$ and then use standard Kalman filter equations for a model with an unknown constant variance parameter to estimate the posterior distributions of $\btheta_{1:T}$ and $\sigma^2$ \citep[Section 4.3.2]{prado2010time}.
The posterior distributions of $\btheta_t$ and $\sigma^2$ after each time step $t$ can be expressed as:
\begin{align*}
    p(\btheta_t \mid \sigma^2, w, \blambda, \bpsi_{1:t}) &\sim N(\btheta_t \mid \mathbf{m}_t, \sigma^2 V_t) \\
    p(\sigma^2 \mid w, \blambda, \bpsi_{1:t}) &\sim \text{Inv-Gamma}(\sigma^2 \mid a_t, b_t),
\end{align*}
where $\mathbf{m}_t$, $V_t$ $a_t$, and $b_t$ are computed using the transformed data as:
\begin{align}
	V_t &= ( (V_{t-1} + wI_p)^{-1} + \bar{X}_t^T \bar{X}_t)^{-1} \nonumber\\ 
	\mathbf{m}_t &= V_t ((V_{t-1} + wI_p)^{-1} \mathbf{m}_{t-1} + \bar{X}_t^T \bpsi_t)  \nonumber\\ 
	a_t &= a_{t-1} + \frac{1}{2} n_t \nonumber\\ 
	b_t &= b_{t-1} + \frac{1}{2}[\mathbf{m}_{t-1}^T (V_{t-1} + wI_p)^{-1} \mathbf{m}_{t-1} + \bpsi_t^T \bpsi_t - \mathbf{m}_t^T V_t^{-1} \mathbf{m}_t] \nonumber \\
	&= b_{t-1} + \frac{1}{2}(\bpsi_t - \bar{X}_t \mathbf{m}_{t-1})^T (I_{n_t} + \bar{X}_t (V_{t-1} + wI_p) \bar{X}_t^T)^{-1} (\bpsi_t - \bar{X}_t \mathbf{m}_{t-1}), \nonumber 
\end{align}
where we initialize $V_0 = v_0 \cdot I_p$ and $\mathbf{m}_0 = \mathbf{0}$.

It is worth noting
that the $V_t$ matrix allows the variance of $\btheta_t$ to vary from athlete to athlete as a function of sample size.
Athletes with many observed scores in a particular rating period will have less posterior uncertainty around their latent ability parameter for that period.
The observation variance parameter $\sigma^2$, on the other hand, is driven by residual variation across all athletes, as demonstrated by the update equation for $b_t$.
Finally, we note that under this framework, if an athlete does not compete in a given time period, the estimated mean of their latent ability parameter does not change.
The variance of their latent ability parameter, however, still increases by $w \cdot \sigma^2$ to reflect the increased uncertainty about their current ability.

To estimate $w$ and $\blambda$ for step~1, we use the marginal posterior density of $w$ and $\blambda$ where $\btheta_{1:T}$ and $\sigma^2$ are integrated out,
which can be expressed in closed form (up to a normalizing constant) as:
\begin{align} \label{eq:objective}
p(w, \blambda &\mid \by_{1:T_{\text{train}}})
    \propto J(\bpsi_{1:T_{\text{train}}} \to \by_{1:T_{\text{train}}}) p(w) p(\blambda)
		\prod_{t=1}^{T_{\text{train}}} p(\bpsi_t \mid \bpsi_{1:t-1}, w, \blambda) \nonumber
\end{align}
for the priors on $w$ and $\blambda$ and posterior predictive densities:
\begin{equation}
	p(\bpsi_{t} \mid \bpsi_{1:t-1}, w, \blambda) 
    = t_{2a_{t-1}}(\bpsi_t \mid \bar{X}_t \mathbf{m}_{t-1}, \frac{b_{t-1}}{a_{t-1}} [I_{n_t} + \bar{X}_t (V_{t-1} + w I_p) \bar{X}_t^T]), \nonumber
\end{equation}
where the $V_t$, $\mathbf{m}_t$, $a_t$, and $b_t$ values are all computed using the Kalman filter as described above.
We can obtain samples of $w$ and $\blambda$ from $p(w, \blambda \mid \by_{1:T_{\text{train}}})$ using standard Markov Chain Monte Carlo (MCMC) methods.
We implement our model in Stan \citep{rstan}, which uses Hamiltonian Monte Carlo and a no-U-turn sampler \citep{hoffman2014no}.

In practice, we can instead take a maximum a posteriori (MAP) approach using standard optimization routines to greatly reduce the computational burden.
Note that each evaluation of $p(\bpsi_t \mid \bpsi_{1:t-1}, w, \blambda)$ for $t = 1, \dots, T_\text{train}$ in the marginal posterior density $p(w, \blambda \mid \by_{1:T_\text{train}})$ requires a Kalman filter to be run on the full training dataset.
MCMC sampling requires many such likelihood evaluations, so even though each Kalman filter run is relatively efficient, MCMC sampling ultimately requires significant computation.
For most applications, however, we are only interested in estimating reasonable values for $w$ and $\blambda$, not in conducting full Bayesian inference on them; as such, it often makes sense to simply find the posterior mode of $p(w, \blambda \mid \by_{1:T_{\text{train}}})$, which requires fewer evaluations of $p(\bpsi_t \mid \bpsi_{1:t-1}, w, \blambda)$ for $t = 1, \dots, T_\text{train}$.
Importantly, integrating $\btheta_{1:T_\textbf{train}}$ and $\sigma^2$ out of the posterior instead of maximizing $p(w, \blambda, \btheta_{1:T_\text{train}}, \sigma^2 \mid \by_{1:T_{\text{train}}})$ with respect to each parameter produces a better-informed posterior mode of $w$ and $\blambda$.

Any nonlinear optimization algorithm can be used to obtain MAP estimates.
For the constrained optimization (with a Dirichlet prior on $\blambda$), we choose to use the Augmented Lagrangian Adaptive Barrier Minimization Algorithm \citep{alabama}.
For the unconstrained optimization (with normal priors on the $\lambda_b$ parameters), we find that the Nelder-Mead algorithm \citep{nelder1965simplex} generally gives the most stable results, though the L-BFGS algorithm \citep{liu1989limited} is much faster in practice.
While these optimization-based approaches may theoretically get stuck in local modes, we find that in practice they produce reasonable and effective results.

We implement the full model-fitting procedure in the \texttt{dlmt} package in \texttt{R}.
Full implementation details for our model-fitting procedure can be found in Appendix \ref{AppA_imp}, and a simulation study confirming appropriate parameter recovery can be found in Appendix \ref{AppB}.

\subsection{Smoothing}
\label{sec:smooth}

The Kalman filter equations produce estimates of the filtered latent ability parameter distributions $p(\btheta_t \mid \bpsi_{1:t}, \sigma^2, \blambda)$.
In the second stage of the model-fitting procedure, we may also wish to calculate the smoothed latent ability parameter distributions $p(\btheta_t \mid \bpsi_{1:T}, \sigma^2, \blambda)$, where the full dataset informs each latent ability parameter estimate.
The Rauch-Tung-Striebel smoother \citep{rauch1965maximum} provides a simple algorithm for doing so.
We can compute the smoother updates as:
\begin{align}
	p(\btheta_t \mid \bpsi_{1:T}, \sigma^2, \blambda) 
		&= N(\btheta_t \mid \mathbf{m}_t^s, \sigma^2 V_t^s) \nonumber
\end{align}
for:
\begin{align}
	\mathbf{m}_t^s &= \mathbf{m}_{t} + S_t (\mathbf{m}_{t+1}^s - \mathbf{m}_t) \nonumber \\
	V_t^s &= V_t + S_t (V_{t+1}^s - V_t - wI) S_t^T \nonumber
\end{align}
for scaling matrix $S_t = V_t (V_t + wI)^{-1}$, where the $\mathbf{m}_t$ and $V_t$ values are the original $\mathbf{m}_t$ and $V_t$ values computed using the Kalman filter equations.
The smoother updates are computed starting from rating period $t=T$ backwards to rating period $t=1$, starting from $\mathbf{m}_T^s = \mathbf{m}_T$ and $V_T^s = V_T$.

\section{Empirical results}
\label{sec:results}

\subsection{USOPC athletic data}

We illustrate our model on a variety of Olympic sport datasets provided to us by the US Olympic and Paralympic Committee.
The data roughly span from 2004 to 2019 and include the score outcomes from selected national and international competitions.
We briefly describe each dataset below:

\paragraph*{Biathlon}{
The biathlon data come from the men's 20km individual biathlon and the men's $4 \times 7.5$ km relay.
In the 20km biathlon, athletes ski a cross-country track between four rifle-shooting rounds.
In each shooting round, they shoot at five targets.
Each miss incurs a penalty, which may be extra time added or a penalty skiing lap, depending on the particular race's rules.
The biathlon relay is similar, with two shooting rounds per relay leg.
In both competitions, athletes compete to finish the race as quickly as possible, so we use each athlete's total time (in seconds) as their score.
}

\paragraph*{Diving}{
The diving data come from women's 3m springboard.
In diving competitions, athletes receive scores for each of their dives in a round.
After each round of diving, only the top-scoring athletes may qualify for the next round.
Only the scores from the final round of competition are recorded, so we use these scores as the game outcomes.}

\paragraph*{Fencing}{
The fencing data come from women's sabre fencing.
In each bout, the first athlete to score fifteen points wins.
We record the score difference between the two athletes as the outcome of each game.
}

\paragraph*{Rugby}{
The rugby data come from men's rugby sevens.
We record the score difference between the two teams as the outcome of each game.
}

\vspace{4mm}

Table \ref{tab:datasum} summarizes the dimensions of the five datasets.
We see a wide range in the numbers of athletes and events across the five sports.
For example, the biathlon dataset contains over 700 individual athletes who compete in only one or two large multicompetitor events per rating period, with each event averaging over 100 competitors.
At the other end of the spectrum, the rugby dataset contains only 90 national teams that compete in many head-to-head games across the 71 rating periods.
In the individual-competitor sports (i.e., biathlon, diving, fencing), there is significant right skew in the number of events per athlete; most athletes in the data compete in only a couple of events, but some athletes compete in many events.
The teams in biathlon relay and rugby tend to compete more consistently.

For the multicompetitor sports of biathlon, biathlon relay, and diving, we divide the sixteen years of data into biannual (i.e., six-month-long) rating periods, resulting in roughly 30 rating periods.
The head-to-head sports of fencing and rugby naturally have more games, so we divide them into quarterly (i.e., three-month-long) rating periods.\footnote{The fencing data starts in 2010 rather than 2004, so there are only 35 total rating periods.}
Choosing a shorter rating period allows more flexibility for athlete abilities to change, but reduces the number of games that can be used to infer abilities within the rating period.
The results shown in this section are generally robust to the choice of rating period (see Appendix \ref{AppA_ratingperiodsens} for details).

\begin{table}[t]
\centering
\begin{tabular}{|l|l|l|l|l|l|l|l|}
\hline
\rowcolor[HTML]{C0C0C0} 
Sport          & Athletes & $T$ & Events & Avg. $\frac{\text{Athletes}}{\text{Event}}$ & Min. $\frac{\text{Events}}{\text{Athlete}}$ & Med.$\frac{\text{Events}}{\text{Athlete}}$ & Max. $\frac{\text{Events}}{\text{Athlete}}$ \\ \hline
Biathlon       & 703      & 31             & 56     & 106                    & 1                                           & 4    & 49   \\ \hline
Biathlon Relay & 30       & 31             & 80     & 18.6                   & 1                                           & 60   & 80   \\ \hline
Diving         & 459      & 32             & 218    & 15                     & 1                                           & 3    & 80   \\ \hline
Fencing        & 489      & 35             & 5806   & 2                      & 1                                           & 14   & 361  \\ \hline
Rugby          & 90       & 71             & 6639   & 2                      & 5                                           & 57   & 784  \\ \hline
\end{tabular}
\caption{Summary table of sport dataset dimensions, containing number of athletes, number of rating periods, number of events, average number of athletes per event, and minimum/median/maximum number of events per athlete.}
\label{tab:datasum}
\end{table}

\subsection{Model fitting and validation}


We fit our unconstrained model to the biathlon, biathlon relay, diving, fencing, and rugby datasets.
We conduct full MCMC (four chains, 1000 burn-in iterations, 1000 samples) as well as MAP estimation using the Nelder-Mead and L-BFGS optimization algorithms.
To assess the convergence of our MCMC estimates of the posterior distributions of $w$ and $\blambda$, we check trace plots and the $\hat{R}$ diagnostic for Hamiltonian Monte Carlo.
Visual inspection of the trace plots does not suggest any evidence of non-convergence, and $\hat{R}$ is nearly equal to one for all model parameters.
The Nelder-Mead and L-BFGS algorithms converge under default convergence tolerances.

\begin{figure}[t]
    \centering
    \includegraphics[width=5.5in]{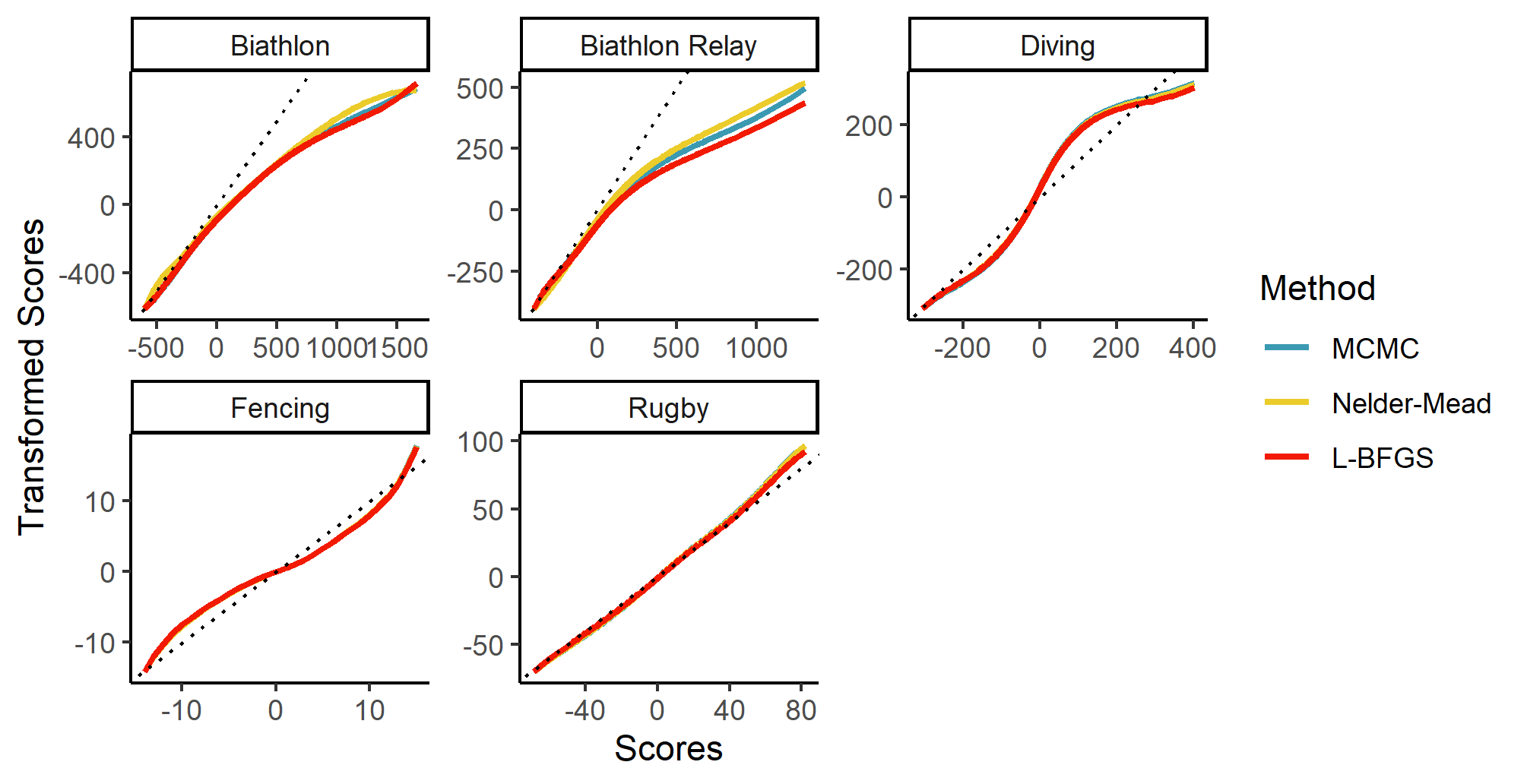}
    \caption{Comparison of algorithms for estimating $\blambda$. 
        Dotted line represents the identity transformation, $y=x$.}
    \label{fig:transformations_comp}
\end{figure}

Before examining the DLM results, we first confirm that the MAP estimates produce similar results as full MCMC.
Figure \ref{fig:transformations_comp} compares the posterior mean of the transformations learned using full MCMC to the transformations learned using MAP with the Nelder-Mead and L-BFGS algorithms.
We see that the learned transformations are essentially the same across all three algorithms for all five sports.
The learned $w$ parameters (not shown) are also very similar.
As a result, using MAP methods rather than full MCMC does not significantly impact model performance, while it can lead to significant speedups.
For example, for the biathlon dataset (\textasciitilde 6000 observations from \textasciitilde 700 athletes over \textasciitilde 60 events), full MCMC took roughly 8 hours, but the Nelder-Mead optimization took only 30 minutes, and L-BFGS converged in less than one minute (all on a laptop with an Intel Core i7-8550U CPU).
To simplify assessment, visualization, and discussion of our results, we will focus on the transformations estimated using full MCMC moving forward, though results are naturally similar for the transformations estimated using MAP.

\begin{figure}[b]
    \centering
    \includegraphics[width=5.5in]{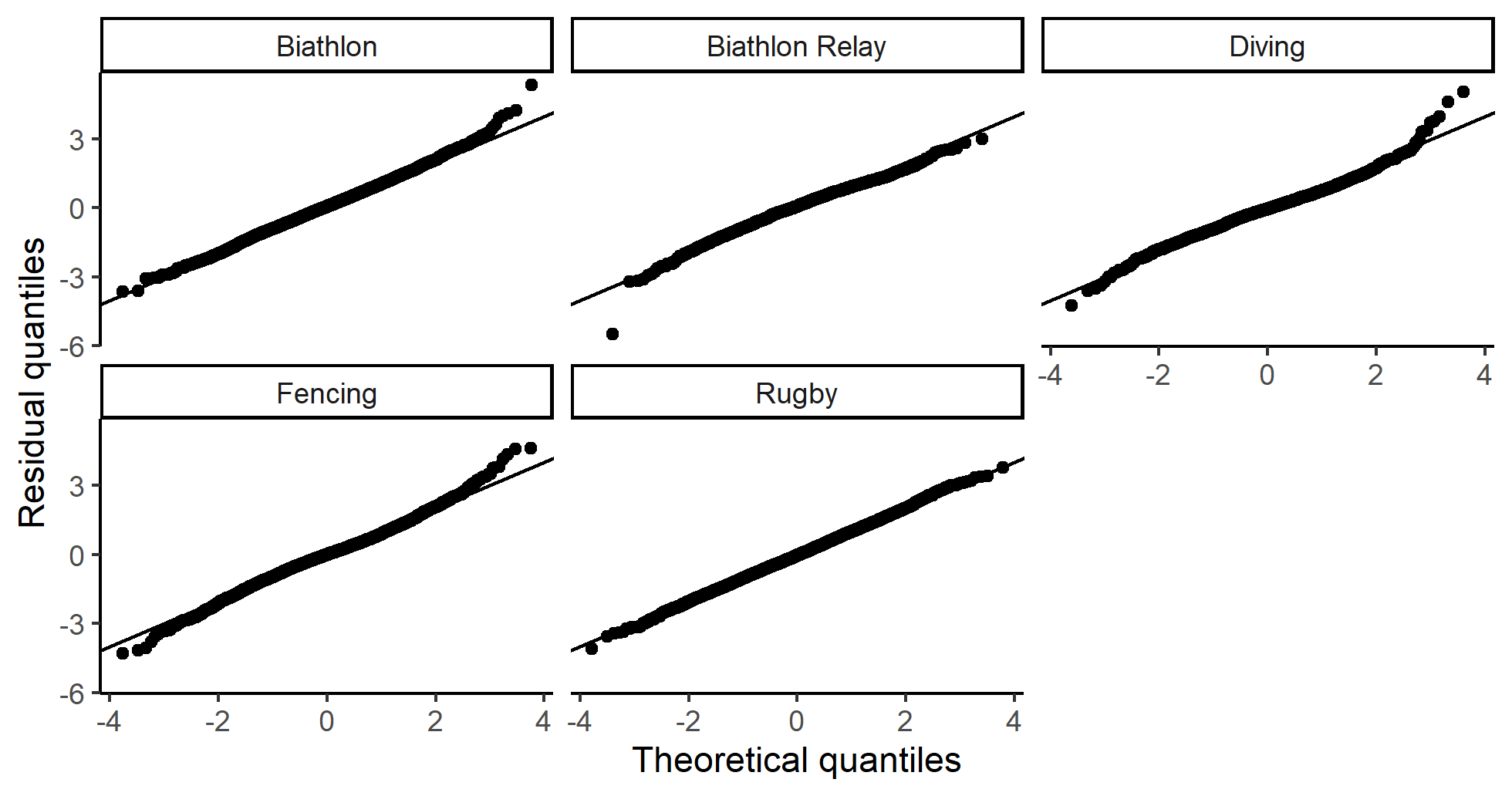}
    \caption{Q-Q plots of test-set residuals compared to standard normal quantiles}
    \label{fig:residuals}
\end{figure}

Next, we assess the fit of the DLM on our transformed datasets.
To do so, we use the first two-third rating periods as a training set to learn an appropriate transformation and leave the last one-third rating periods as a test set.
We then visualize $(\bpsi_t - \bar{X}_t \mathbf{m}_{t-1})$, i.e., the one-step prediction residuals, on the test set.
If the learned transformation is effective, the residuals should be approximately normally distributed.
Figure \ref{fig:residuals} shows Q-Q plots of standardized test-set residuals against standard normal quantiles.
While the residuals show some slight outliers at the extremes, they are largely normally distributed.
The learned transformations thus appear to produce reasonable fits of the standard DLM to these datasets.


\subsection{Case studies: biathlon and rugby}

We use the biathlon and rugby datasets to illustrate the results from our model.
For both datasets, we learn an unconstrained monotone spline transformation using MCMC, transform the dataset, and run the Kalman filter on the transformed data.

\begin{figure}[t]
    \centering
    \includegraphics[width=5.5in]{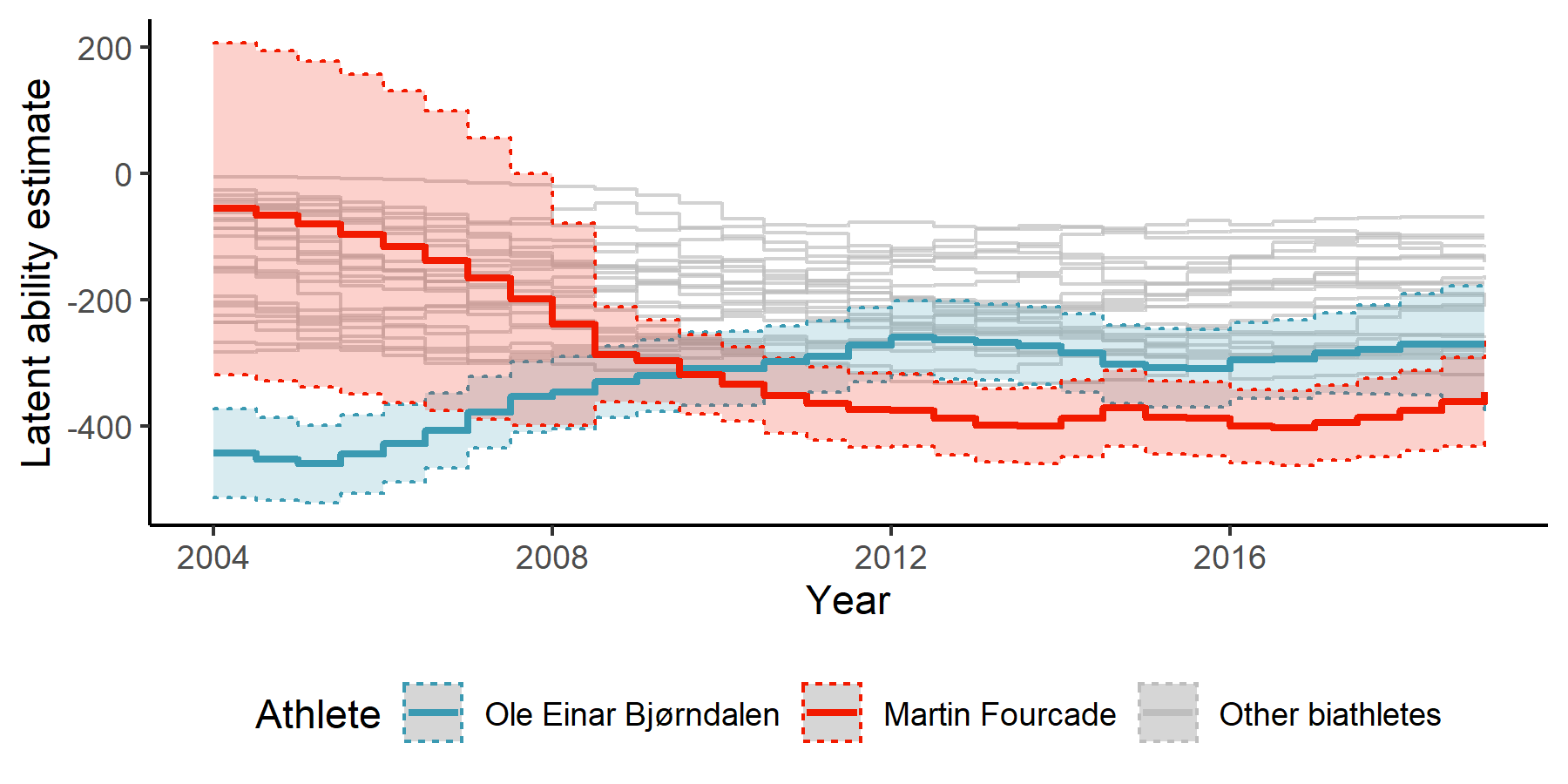}
    \caption{Means from 2004-2018 of ability parameters of 25 biathletes with the most races entered, estimated using the Rauch-Tung-Striebel smoother (Section \ref{sec:smooth}). 90\% central posterior interval shown for Ole Einar Bjørndalen and Martin Fourcade.}
    \label{fig:ole}
\end{figure}

Figure \ref{fig:ole} shows the smoothed biannual latent ability point estimates for the 25 biathletes with the most biathlons entered.
In the biathlon, low race times are better, so negative ability parameters indicate strong athletes.
Of the visualized latent ability trajectories, two stand out, and Figure \ref{fig:ole} additionally shows their 90\% central posterior probability intervals.
The trajectory in blue belongs to the ``King of Biathlon,'' Ole Einar Bjørndalen, the winningest biathlete of all time at the Olympics, Biathlon World Championships, and the Biathlon World Cup tour.
Though the scope of our dataset does not include the start of his career in the 1990s, the model clearly notes his dominance in the early 2000s.
The trajectory in red belongs to Martin Fourcade, who began serious international competition in 2008 (hence the wide probability intervals prior to 2008) and proceeded to put together a record-breaking string of seven overall World Cup titles in a row from 2011 to 2018.
Bjørndalen and Fourcade are the two names considered in discussions of the greatest male biathlete of all time.
Our model notes Fourcade's quick rise to prominence, projecting that he would begin to outperform Bjørndalen as early as 2009, though interestingly it never projects Fourcade's latent ability to exceed Bjørndalen's peak latent ability in 2005.

\begin{figure}[t]
    \centering
    \includegraphics[width=5.5in]{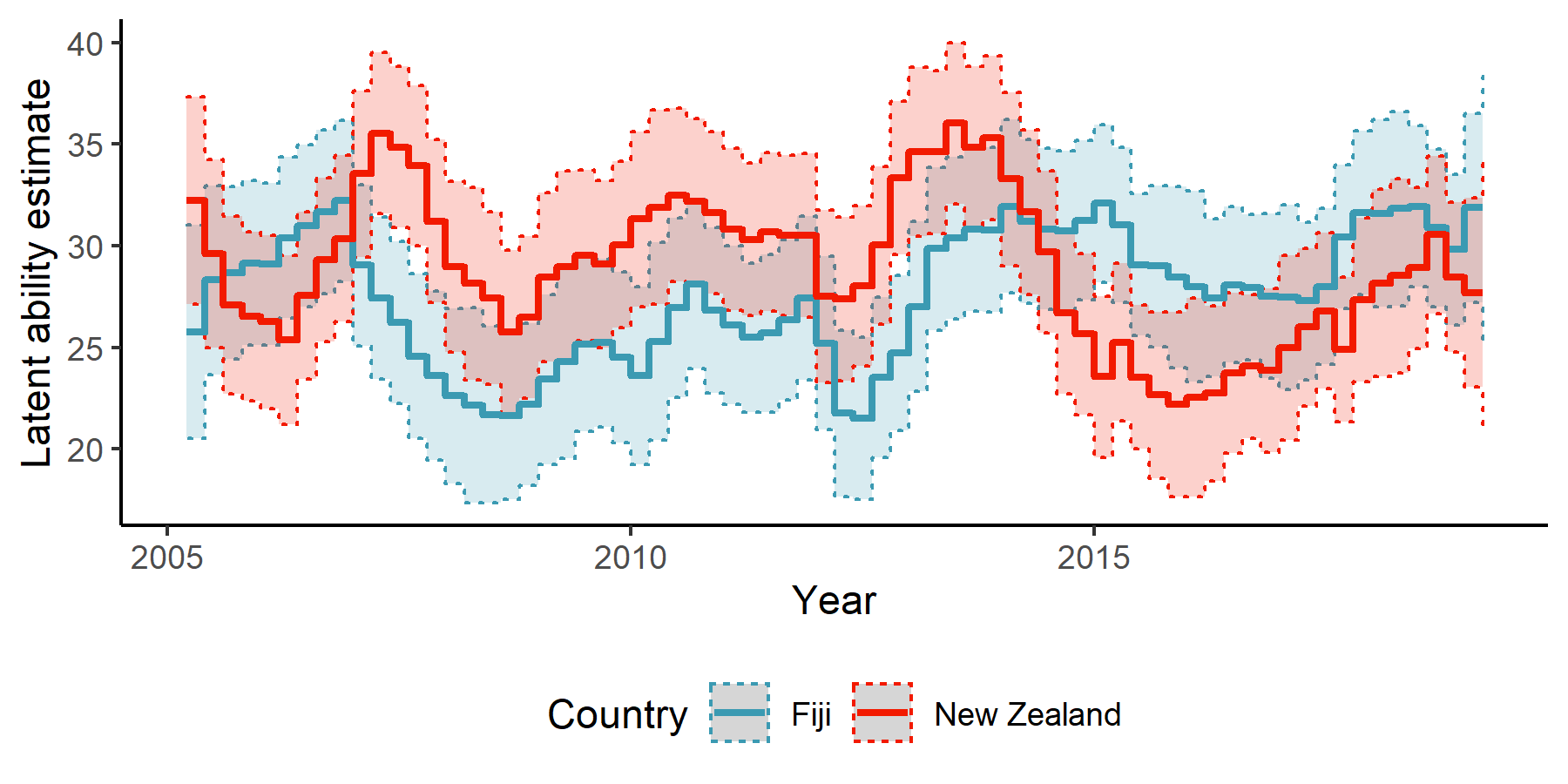}
    \caption{Means of ability parameters and 90\% central posterior intervals shown for Fiji and New Zealand national teams, estimated using the Rauch-Tung-Striebel smoother (Section \ref{sec:smooth})}.
    \label{fig:rugby}
\end{figure}

Figure \ref{fig:rugby} shows the smoothed quarterly latent ability point estimates and corresponding 90\% posterior intervals for the national rugby teams of Fiji and New Zealand, two countries popularly known for their strong rugby teams.
Here, more positive latent ability estimates represent stronger teams.
While the two teams have similar estimated strengths during the time span of the data, we see that the model briefly estimates New Zealand to be significantly stronger in 2007 and 2013, and Fiji to be significantly stronger in 2015.


\subsection{Model results}

\begin{figure}[t]
    \centering
    \includegraphics[width=5.5in]{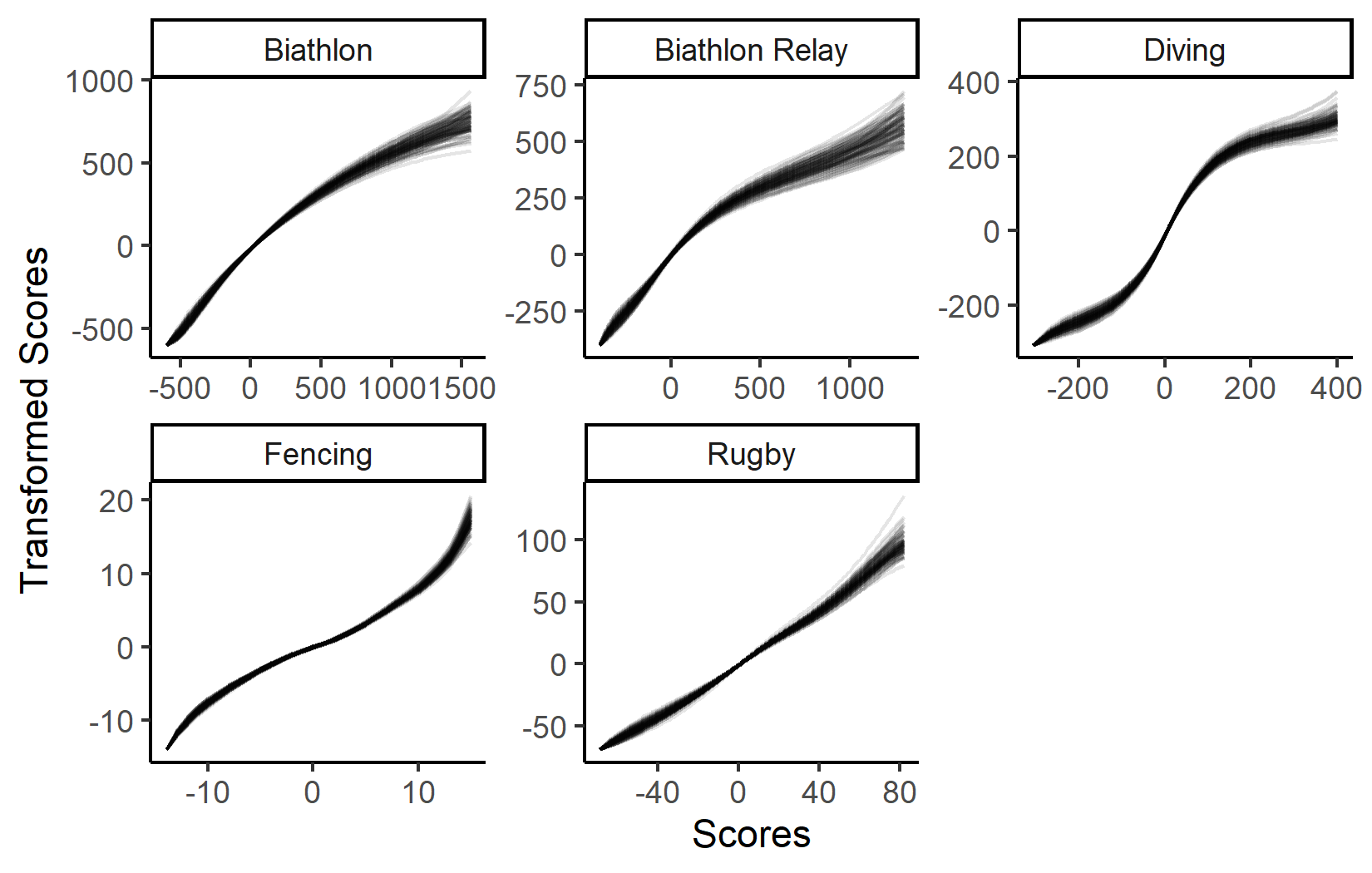}
    \caption{Transformations learned by DLM with transformations}
    \label{fig:transformations}
\end{figure}

Another result of interest is the monotone spline transformation learned by the DLM with transformations.
Figure \ref{fig:transformations} displays 100 posterior transformation samples for each of the five sports, under the unconstrained optimization.
The transformations generally reflect conventional wisdom about scores from these sports.
For example, we sometimes see very slow race times in the biathlon and biathlon relay, which occur when an athlete makes a few shooting mistakes and takes penalties.
We expect to see a negative feedback loop in the biathlon where taking penalties causes athletes to get more frustrated or tired from penalty laps, which causes more penalties.
This means that extremely slow race times may not reflect extremely poor skill.
The learned transformation shrinks the very slow race times to be less extreme, which intuitively helps to make them better reflect athlete skill in the normal DLM.
On the other hand, the transformation learned for fencing magnifies extreme score differences.
In a fencing bout, points are scored one-at-a-time; after each touch (i.e., point scored), the fencers reset to their starting positions.
This makes very one-sided matches relatively rare, since even outmatched fencers can typically score some number of lucky points in a match to fifteen points.
The transformation notes this and indicates that when a fencer wins by many points, they are much stronger than their opponent, even more so than the large score gap may suggest.

Table \ref{tab:wsig} shows the posterior means of the $\sqrt{w}$ and $\sigma$ parameters for each of the five sports.
Recall that $\sigma^2$ represents the observation variance and $w$ represents the ratio of the innovation variance to the observation variance, so $\sigma$ and $\sigma \sqrt{w}$ would represent the observation and innovation standard deviations, respectively.
For example, in the biathlon data, the model estimates that the innovation standard deviation is approximately 28\% as large as the observation standard deviation of 98.5.
We make two notes about the $\sqrt{w}$ values estimated in these datasets.
First, we see that $\sqrt{w} < 1$ in all five datasets, i.e., that the athletes' period-to-period variability in their latent abilities is generally much smaller than the game-to-game variability in their scores.
The competitors in our datasets are all professional-level athletes.
As such, while they may still improve or worsen over time, their game scores are largely driven by how well they perform in each particular game rather than by major changes to their skill level.
Secondly, we note that the $\sqrt{w}$ values vary across sports.
In addition to capturing period-to-period variability, the value of $\sqrt{w}$ also reflects the extent to which skill, rather than random chance, appears to drive game results; 
for example, a game with results driven almost entirely by skill would have minimal game-to-game variation, i.e., $\sigma \approx 0$, so any period-to-period variation in athlete abilities would imply a large value of $\sqrt{w}$.
The results in Table \ref{tab:wsig} therefore suggest that of the five sporting events, the results from diving competitions may be the most driven by skill.
On the other hand, the random chance associated with particular performances appears to play a significant role in the outcomes of fencing bouts, which has been noted by other sources as well (e.g., \citet{zappala2022role}).
\begin{table}[t]
    \centering
    \begin{tabular}{|l|l|l|}
    \hline
    \rowcolor[HTML]{C0C0C0} 
    Sport          & $\sqrt{w}$ & $\sigma$ \\ \hline
    Biathlon       & 0.28 & 98.5 \\ \hline
    Biathlon relay & 0.22 & 71.6 \\ \hline
    Diving         & 0.40 & 49.7 \\ \hline
    Fencing        & 0.06 & 3.2  \\ \hline
    Rugby          & 0.18 & 14.5 \\ \hline
    \end{tabular}
    \caption{Posterior means of variance parameters. 
    Note that the $\sigma$ parameters are on the scale of the transformed data rather than the original scale of the data.}
    \label{tab:wsig}
\end{table}

\subsection{Comparing rating methods}

To evaluate the DLM with transformations, we compared the accuracy of its predictions to predictions made by other models for multi-competitor and head-to-head athlete rating.
For multi-competitor sports, we compared the DLM with transformations (LM-T) to the DLM without transformations (LM) and the dynamic rank-order logit model (ROL) from \citet{glickman2015stochastic}.
For head-to-head sports, we compare to the Glicko rating system \citep[GLO;][]{glickman1999parameter}.
We use the first two-third rating periods in each dataset as a training set to tune the model hyperparameters $w$ and $\blambda$, fit the model on the full dataset, and finally evaluate its predictions for the test set (i.e., the last one-third rating periods).
For multi-competitor games, we evaluate predictions using the Spearman correlations between the observed and predicted athlete rankings in each game.
This approach was taken in \citet{glickman2015stochastic} to evaluate predictability on a test set.
We summarize these game correlations $\rho_{tg}$ over the test set using a game-size-weighted average \citep{glickman2015stochastic}:
\begin{equation}
	\rho = \frac{\sum_{t=\lceil\frac{2}{3}T\rceil}^{T} \sum_{g=1}^{g_t} (n_{tg} - 1) \rho_{gt}}{\sum_{t=\lceil\frac{2}{3}T\rceil}^{T} \sum_{g=1}^{g_t} (n_{tg} - 1)}. \label{eq:wtcor}
\end{equation} 
Note that we limit ourselves to using rank-based metrics to facilitate comparison with the ROL model, which predicts athlete ranking probabilities rather than scores.
For head-to-head games, we evaluate predictions using the average accuracy of winner predictions in the test set.

Tables \ref{tab:multi} and \ref{tab:h2h} show the weighted Spearman correlation (Equation \ref{eq:wtcor}) of the ranking predictions for the multi-competitor sports and the accuracy of winner/loser predictions for the head-to-head sports.
\begin{table}[t]
\centering
\begin{tabular}{|c|ccc|}
    \hline
    \multicolumn{1}{|l|}{}                 & \multicolumn{3}{c|}{\cellcolor[HTML]{EC8F9C}Model} \\ \hline
    \rowcolor[HTML]{E9E5DC} 
    \cellcolor[HTML]{EC8F9C}Sport          & \multicolumn{1}{c|}{\cellcolor[HTML]{E9E5DC}LM-T} & \multicolumn{1}{c|}{\cellcolor[HTML]{E9E5DC}LM} & ROL  \\ \hline
    \cellcolor[HTML]{E9E5DC}Biathlon & 
        \multicolumn{1}{c|}{.64} & \multicolumn{1}{c|}{.61} & .61 \\ \hline
    \cellcolor[HTML]{E9E5DC}Biathlon Relay & 
        \multicolumn{1}{c|}{.77} & \multicolumn{1}{c|}{.75} & .75 \\ \hline
    \cellcolor[HTML]{E9E5DC}Diving & 
        \multicolumn{1}{c|}{.64} & \multicolumn{1}{c|}{.62} & .61 \\ \hline
\end{tabular}
    \caption{Weighted spearman correlations of predictions for the 18 biathlon events, 27 biathlon relay events, and 83 diving events in the test sets.}
    \label{tab:multi}
\end{table}
\begin{table}[t]
\centering
\begin{tabular}{|c|ccc|}
    \hline
    \multicolumn{1}{|l|}{}          & \multicolumn{3}{c|}{\cellcolor[HTML]{EC8F9C}Model}                                                         \\ \hline
    \rowcolor[HTML]{E9E5DC} 
    \cellcolor[HTML]{EC8F9C}Sport   & \multicolumn{1}{c|}{\cellcolor[HTML]{E9E5DC}LM-T} & \multicolumn{1}{c|}{\cellcolor[HTML]{E9E5DC}LM} & GLO  \\ \hline
    \cellcolor[HTML]{E9E5DC}Fencing & 
        \multicolumn{1}{c|}{.70} & \multicolumn{1}{c|}{.67} & .68 \\ \hline
    \cellcolor[HTML]{E9E5DC}Rugby   & 
        \multicolumn{1}{c|}{.71} & \multicolumn{1}{c|}{.72} & .70  \\ \hline
\end{tabular}
\caption{Accuracy of winner predictions \label{tab:h2h} across the 1,785 fencing events and 2,503 rugby events in the test set.}
\end{table}
Across the five datasets, we see some evidence that the LM-T model outperforms the other models in terms of predictive performance.
While it is difficult to generally evaluate the relative empirical performance of the LM-T model with only five datasets, we see that it improves predictive accuracy across nearly all of the datasets.
The only exception is the rugby dataset, where the LM-T model essentially learns an identity transformation, so we would not expect it to outperform the LM model.
The results of athletic competitions are generally challenging to predict; small improvements in predictive performance, such as those demonstrated above, can therefore be fairly valuable for generating a competitive edge.

\section{Discussion}

In this paper, we introduce a novel model to rate athletes who compete in head-to-head and multi-competitor sports with score outcomes.
Using observed scores rather than rankings to rate athletes provides additional information, which generally improves predictions in the settings we consider.
We can fit the model either using MCMC or an MAP approach, both which utilize the computational efficiency of the Kalman filter to learn an appropriate transformation to apply to the score outcomes.
The full model-fitting procedure is implemented in the \texttt{dlmt} package in \texttt{R}.


The simple normal DLM at the core of our model makes a variety of extensions possible.
For example, we choose to use a normal random walk as the innovation process for athletes' latent abilities, which may easily be replaced by alternative innovation processes, such as a mean-preserving random walk \citep{glickman2015stochastic} or an autoregressive process \citep{glickman1998state}.
Also, external covariates related to athletes (e.g., height, age, experience), events (e.g., weather conditions), and/or other factors may be assigned fixed or time-varying coefficients and straightforwardly incorporated into the normal likelihood and innovation equations.
If transforming outcomes to normality is infeasible in a particular setting, the normal likelihood could be also extended to the likelihood of a generalized linear model \citep{west1985dynamic}.

The model introduced in this paper assumes a constant observation variance parameter $\sigma^2$ for computational efficiency and conceptual simplicity.
The $V_t$ matrices enable the variances of athletes' latent ability parameters to differ based on the number of games they have played, reflecting uncertainty in the estimation of their latent abilities.
They do not, however, allow two athletes who have played the same number of games in each rating period to have different variances, which would reflect fundamental differences in performance variation across athletes.
Modeling this type of heteroskedasticity across athletes may be useful, particularly in data-rich settings (e.g., settings where most athletes play many games) where it would be feasible to estimate athlete-specific variance parameters.
Extending the model to learn athlete-specific heteroskedasticity parameters while preserving efficient closed-form Kalman filter updates is beyond the scope of our work and a direction for future research.


While we focus on athlete rating, our model may be used for a wide range of different problems.
Dynamic linear models are very popular for analyzing time series data in fields ranging from engineering and finance to health and ecology.
In many of these applications, the normal likelihood in a standard normal DLM may be misspecified, which can be addressed by learning an order-preserving monotone transformation using the model introduced in this paper.

The problem of athlete rating has interested organizations and individuals alike for many years.
Appropriately using the information contained in game scores is an important but challenging task, due to the unusual features of score information in different games.
This paper provides a general method to address these challenges, which can be applied to a wide range of multi-competitor and head-to-head games.

\section*{Acknowledgements}

We thank Dan Webb at the U.S. Olympic and Paralympic Committee for providing the data for this work.
We also thank our reviewers for their helpful comments.
This research was supported in part by a research contract from the U.S. Olympic and Paralympic Committee, and by the National Science Foundation Graduate Research Fellowship Program under Grant No. DGE1745303.

\bibliographystyle{imsart-nameyear} 
\bibliography{refs.bib}       

\appendix
\section{Model implementation details}

\subsection{Practical implementation details}
\label{AppA_imp}

We make a few minor changes to the model presented in section \ref{sec:fitting}.
We clear the off-diagonal elements of the $V_t$ matrix between rating periods in the Kalman filter equations.
This keeps the matrix operations relatively sparse, which speeds up computation time.
The correlations between players induced by playing in the same games tend to be weak in any case, so this change does not hurt model performance.
We also cap the diagonal entries of $V_t$ at $v_0$, assuming that we cannot be less certain about a player's ability than we are before seeing any data.
Appendix \ref{AppB} contains a simulation study confirming that these approximations do not significantly hurt model performance.


In this paper, we consider splines of degree three, with three knots placed at the $25^\text{th}$, $50^\text{th}$, and $75^\text{th}$ percentiles of the raw centered data.
In experiments (not shown here) with the datasets analyzed in this paper, we found that three knots generally provided sufficient flexibility such that the transformations estimated with four or more knots did not produce substantially different transformations from those estimated with three knots.
Increasing the knots past six typically led to unstable optimization results and unusual transformations with poor predictive power for our datasets.
We suggest trying a few different placements of three knots to determine sensitivity to knot placement.
In particular, some applications may require significant flexibility in the tail regions of the transformation, so more knots may need to be placed at extreme quantiles.
Knot choices may be evaluated either by visual inspection of the learned transformation for unusual or unexpected nonlinearities, or simply by evaluation of the resulting predictions on a test set using a scale-free outcome metric.

We implement the monotone spline transformations using the \texttt{splines2} package in \texttt{R} \citep{splines2-paper}.
We implement the Bayesian model using the \texttt{rstan} package in \texttt{R} \citep{rstan}, which uses Hamiltonian Monte Carlo to generate posterior samples.
For MAP estimates, we primarily use the Nelder-Mead algorithm \citep{nelder1965simplex} as implemented in the \texttt{optim()} function in \texttt{R}.
The L-BFGS algorithm \citep{liu1989limited} is implemented via the \texttt{optimizing()} function in the \texttt{rstan} package.

\subsection{Sensitivity to choice of rating period}
\label{AppA_ratingperiodsens}

\begin{figure}[ht]
    \centering
    \includegraphics[width=5.5in]{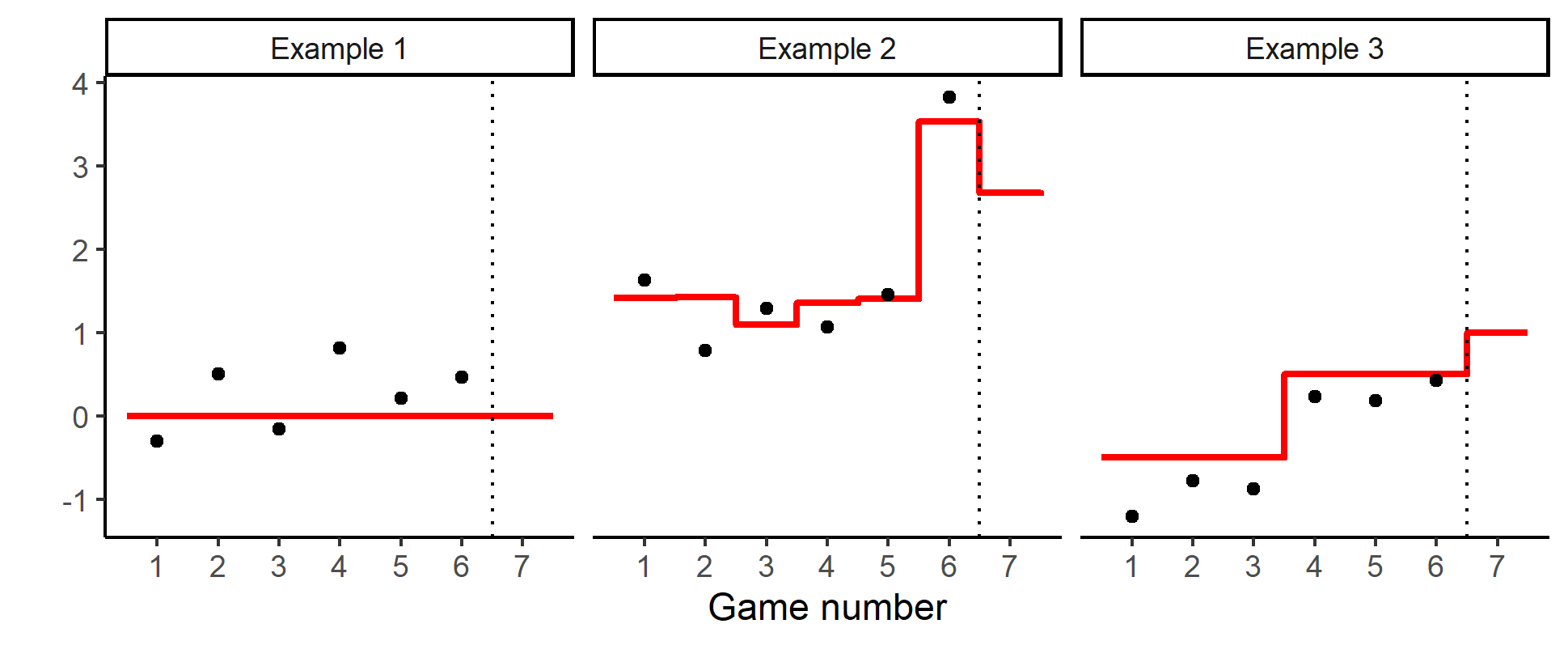}
    \caption{Toy example illustrating bias-variance tradeoff in rating period length.}
    \label{fig:rpsens_toyexample}
\end{figure}

The choice of rating period length reflects a bias-variance tradeoff; as rating period length increases, we subject our estimates to more potential bias, but decrease their variance.
Using different rating periods has minimal effects on our models for the datasets we consider in the paper, as we show at the end of this Appendix.
Nonetheless, we will first show a toy example to illustrate how the bias-variance tradeoff generally occurs.
Figure \ref{fig:rpsens_toyexample} plots three examples of a single athlete's performances across six games.
The athlete's latent ability is plotted as the red curve, and the observed performances are plotted as points.

Consider the problem of predicting the athlete's score $y$ in game seven.
As analysts, we can divide the six previous games into any number of rating periods.
Our best estimate of $y$ would then be the inferred distribution of $\theta$ after game six.
The lowest-variance estimate of $\theta$ would come from grouping all six games into a single rating period, since doing so avoids adding any between-rating-period innovation variance into our estimate of $\theta$.
Such an estimate would be ideal for Example 1, where the athlete's latent ability remains constant over time.
Importantly, using scores from games 1-5 does not incur any bias in Example 1, since these scores are all generated by the same latent ability parameter as is game six.

The lowest-bias estimate of $\theta$, on the other hand, comes from assigning each game its own rating period.
Doing so effectively upweights more recent outcomes, which tend to be the most predictive of future outcomes.
This approach would provide the best results in Example 2, where the athlete's latent ability changes after every game according to a normal random walk.
Adding innovation variance after each game allows the estimated latent ability to strongly adapt to each new data point
In Example 2, this means that the final latent ability estimate will be closer to the result from game six, and won't be biased as much by the observed outcomes from games 1-5, which were generated under different latent ability parameters.

In practice, we recommend navigating this bias-variance tradeoff by choosing rating periods that reflect domain knowledge about how frequently athletes' abilities may experience significant changes.
To predict an athlete's score in a new game, the analyst would ideally rely most heavily on all prior observations generated under the athlete's most recent latent ability parameter.
In Example 3, this would be achieved by using three-game-long rating periods; using shorter rating periods would reduce the precision of the final latent ability estimate, and using longer rating periods would introduce additional bias.
While analysts do not know how frequently athletes' latent ability parameters experience significant changes, the seasonal structure of most sports provides a reasonable framework for deciding rating periods.
For example, athletes' abilities will likely change between seasons due to offseason training.
Longer sporting seasons may exhibit a ``play oneself into shape'' effect, which would require shorter rating periods to account for significant within-season ability changes.

To study the sensitivity of our results to choice of rating period, we run our models on each sport using a variety of reasonable rating periods.
For biathlon, biathlon relay, and diving, which generally have fewer events, we consider annual and biannual rating periods.
We consider annual, biannual, and quarterly rating periods for fencing, and we consider annual, biannual, quarterly, bimonthly, and monthly rating periods for rugby.
We keep the same events in the test set regardless of the choice of rating period, and estimate $w$ and $\blambda$ on the training set using MAP (so results here differ slightly from the results shown the main text).

\begin{figure}[t]
    \centering
    \includegraphics[width=5.5in]{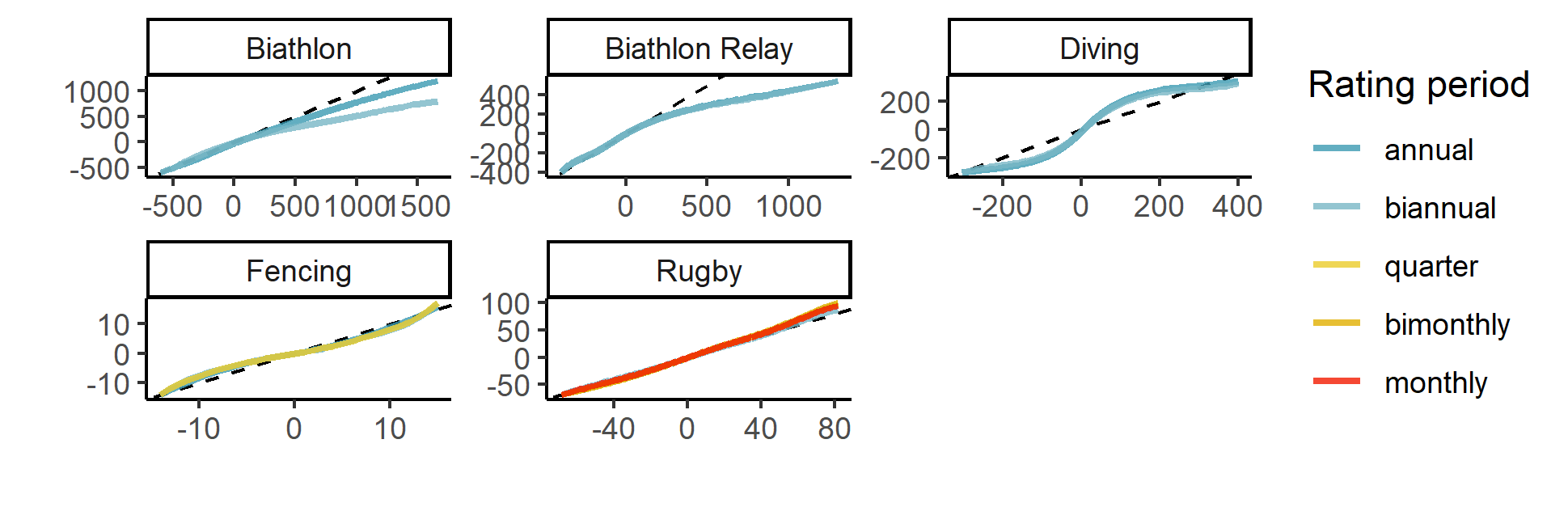}
    \caption{Learned transformations for each sport, using different rating period lengths.}
    \label{fig:sens_trans}
\end{figure}

The primary results of our analyses remain relatively constant across different choices of rating period.
Figure \ref{fig:sens_trans} shows the transformations learned under the different rating period lengths.
We see that using different rating periods has very little effect on the optimal transformations for the datasets we consider.
As a result, using different rating periods also has little effect on the posterior means of the observation standard deviation, as seen in Table \ref{tab:senssig}.
The estimated innovation standard deviation $\sqrt{w}$, on the other hand, naturally decreases as the rating period grows shorter, since there is less time between rating periods for athletes' latent abilities to change.

\begin{table}[t]
\centering
\begin{tabular}{c|ccccc|}
\cline{2-6}
                                                             & \multicolumn{5}{c|}{\cellcolor[HTML]{EC8F9C}Rating Period}                                                                                                                                                                              \\ \hline
\rowcolor[HTML]{E9E5DC} 
\multicolumn{1}{|c|}{\cellcolor[HTML]{EC8F9C}Sport}          & \multicolumn{1}{c|}{\cellcolor[HTML]{E9E5DC}Annual} & \multicolumn{1}{c|}{\cellcolor[HTML]{E9E5DC}Biannual} & \multicolumn{1}{c|}{\cellcolor[HTML]{E9E5DC}Quarterly} & \multicolumn{1}{c|}{\cellcolor[HTML]{E9E5DC}Bimonthly} & Monthly \\ \hline
\multicolumn{1}{|c|}{\cellcolor[HTML]{E9E5DC}Biathlon}       & \multicolumn{1}{c|}{98.0}                           & \multicolumn{1}{c|}{96.7}                             & \multicolumn{1}{c|}{--}                                & \multicolumn{1}{c|}{--}                                & --      \\ \hline
\multicolumn{1}{|c|}{\cellcolor[HTML]{E9E5DC}Biathlon Relay} & \multicolumn{1}{c|}{69.1}                           & \multicolumn{1}{c|}{69.5}                             & \multicolumn{1}{c|}{--}                                & \multicolumn{1}{c|}{--}                                & --      \\ \hline
\multicolumn{1}{|c|}{\cellcolor[HTML]{E9E5DC}Diving}         & \multicolumn{1}{c|}{48.6}                           & \multicolumn{1}{c|}{49.0}                             & \multicolumn{1}{c|}{--}                                & \multicolumn{1}{c|}{--}                                & --      \\ \hline
\multicolumn{1}{|c|}{\cellcolor[HTML]{E9E5DC}Fencing}        & \multicolumn{1}{c|}{3.11}                           & \multicolumn{1}{c|}{3.11}                             & \multicolumn{1}{c|}{3.11}                              & \multicolumn{1}{c|}{--}                                & --      \\ \hline
\multicolumn{1}{|c|}{\cellcolor[HTML]{E9E5DC}Rugby}          & \multicolumn{1}{c|}{14.4}                           & \multicolumn{1}{c|}{14.4}                             & \multicolumn{1}{c|}{14.2}                              & \multicolumn{1}{c|}{13.4}                              & 14.3    \\ \hline
\end{tabular}
\caption{Posterior means of $\sigma$ parameter, using different rating periods.}
\label{tab:senssig}
\end{table}

\begin{table}[t]
\centering
\begin{tabular}{c|ccccc|}
\cline{2-6}
                                                             & \multicolumn{5}{c|}{\cellcolor[HTML]{EC8F9C}Rating Period}                                                                                                                                                                              \\ \hline
\rowcolor[HTML]{E9E5DC} 
\multicolumn{1}{|c|}{\cellcolor[HTML]{EC8F9C}Sport}          & \multicolumn{1}{c|}{\cellcolor[HTML]{E9E5DC}Annual} & \multicolumn{1}{c|}{\cellcolor[HTML]{E9E5DC}Biannual} & \multicolumn{1}{c|}{\cellcolor[HTML]{E9E5DC}Quarterly} & \multicolumn{1}{c|}{\cellcolor[HTML]{E9E5DC}Bimonthly} & Monthly \\ \hline
\multicolumn{1}{|c|}{\cellcolor[HTML]{E9E5DC}Biathlon}       & \multicolumn{1}{c|}{.39}                            & \multicolumn{1}{c|}{.28}                              & \multicolumn{1}{c|}{--}                                & \multicolumn{1}{c|}{--}                                & --      \\ \hline
\multicolumn{1}{|c|}{\cellcolor[HTML]{E9E5DC}Biathlon Relay} & \multicolumn{1}{c|}{.30}                            & \multicolumn{1}{c|}{.21}                              & \multicolumn{1}{c|}{--}                                & \multicolumn{1}{c|}{--}                                & --      \\ \hline
\multicolumn{1}{|c|}{\cellcolor[HTML]{E9E5DC}Diving}         & \multicolumn{1}{c|}{.43}                            & \multicolumn{1}{c|}{.39}                              & \multicolumn{1}{c|}{--}                                & \multicolumn{1}{c|}{--}                                & --      \\ \hline
\multicolumn{1}{|c|}{\cellcolor[HTML]{E9E5DC}Fencing}        & \multicolumn{1}{c|}{.13}                            & \multicolumn{1}{c|}{.10}                              & \multicolumn{1}{c|}{.08}                               & \multicolumn{1}{c|}{--}                                & --      \\ \hline
\multicolumn{1}{|c|}{\cellcolor[HTML]{E9E5DC}Rugby}          & \multicolumn{1}{c|}{.23}                            & \multicolumn{1}{c|}{.17}                              & \multicolumn{1}{c|}{.17}                               & \multicolumn{1}{c|}{.15}                               & .10     \\ \hline
\end{tabular}
\caption{MAP estimates of $w$ parameter, using different rating periods.}
\label{tab:sensw}
\end{table}

Though rating period has little overall effect, we see some evidence of the bias-variance tradeoff as we shorten the rating period.
Tables \ref{tab:sens_multi} and \ref{tab:sens_h2h} show performance of the models on the test set under different rating periods.
As the rating periods decrease in length, the models' predictions experience slightly less bias, which in turn results in slight improvements in predictive performance.
On the other hand, shorter rating periods lead to slightly greater variance in the estimates of athletes' $\theta$ parameters.
Recall that the posterior variance of $\btheta_t$ equals $\sigma^2 V_t$ after each rating period $t$, so the diagonal entries of $V_t$ indicate how the variance in each athlete's $\theta_t$ parameter relates to $\sigma^2$.
To compare athlete variances across different choices of rating period, we record the diagonal entries of $V_t$ for $t = 1, \dots, T$ that correspond to the end of each annual rating period.\footnote{E.g., for a biannual rating period, we would record the diagonal entries of $V_2, V_4, \dots$, which correspond to the ends of annual rating periods $1, 2, \dots$.}
We then average these values over all athletes, over all rating periods to produce an average value for the entries in the $V_t$ matrices given a choice of rating period.
By definition, shorter rating periods always result in larger diagonal entries in $V_t$ at the end of each annual rating period, since they add more innovation variance while maintaining the same total number of events in the annual rating period.
As Table \ref{tab:sensV} shows, however, the average increases in standard deviation are generally quite small across all of the datasets we consider.
For example, in the biathlon relay, changing from annual to biannual rating periods only increases the average $\sqrt{V_t}$ value by about $0.02\sigma$.
Of course, depending on their pattern of games played, particular athletes may see smaller or larger increases in the standard deviation of their $\theta_t$ estimates as the timing window changes;
for example, the largest individual change for an athlete in the diving dataset is 0.52$\sigma$.

\begin{table}[t]
\centering
\begin{tabular}{c|cc|}
\cline{2-3}
                                                             & \multicolumn{2}{c|}{\cellcolor[HTML]{EC8F9C}Rating Period}     \\ \hline
\rowcolor[HTML]{E9E5DC} 
\multicolumn{1}{|c|}{\cellcolor[HTML]{EC8F9C}Sport}          & \multicolumn{1}{c|}{\cellcolor[HTML]{E9E5DC}Annual} & Biannual \\ \hline
\multicolumn{1}{|c|}{\cellcolor[HTML]{E9E5DC}Biathlon}       & \multicolumn{1}{c|}{.63}                            & .64      \\ \hline
\multicolumn{1}{|c|}{\cellcolor[HTML]{E9E5DC}Biathlon Relay} & \multicolumn{1}{c|}{.77}                            & .78      \\ \hline
\multicolumn{1}{|c|}{\cellcolor[HTML]{E9E5DC}Diving}         & \multicolumn{1}{c|}{.61}                            & .64      \\ \hline
\end{tabular}
\caption{Weighted spearman correlations of predictions for events in test set, using different rating periods.}
    \label{tab:sens_multi}
\end{table}
\begin{table}[t]
\centering
\begin{tabular}{c|ccccc|}
\cline{2-6}
                                                      & \multicolumn{5}{c|}{\cellcolor[HTML]{EC8F9C}Rating Period}                                                                                                                                                                              \\ \hline
\rowcolor[HTML]{E9E5DC} 
\multicolumn{1}{|c|}{\cellcolor[HTML]{EC8F9C}Sport}   & \multicolumn{1}{c|}{\cellcolor[HTML]{E9E5DC}Annual} & \multicolumn{1}{c|}{\cellcolor[HTML]{E9E5DC}Biannual} & \multicolumn{1}{c|}{\cellcolor[HTML]{E9E5DC}Quarterly} & \multicolumn{1}{c|}{\cellcolor[HTML]{E9E5DC}Bimonthly} & Monthly \\ \hline
\multicolumn{1}{|c|}{\cellcolor[HTML]{E9E5DC}Fencing} & \multicolumn{1}{c|}{.70}                            & \multicolumn{1}{c|}{.70}                              & \multicolumn{1}{c|}{.71}                               & \multicolumn{1}{c|}{--}                                & --      \\ \hline
\multicolumn{1}{|c|}{\cellcolor[HTML]{E9E5DC}Rugby}   & \multicolumn{1}{c|}{.70}                            & \multicolumn{1}{c|}{.70}                              & \multicolumn{1}{c|}{.71}                               & \multicolumn{1}{c|}{.71}                               & .71     \\ \hline
\end{tabular}
\caption{Accuracy of predictions for events in test set, using different rating periods.}
    \label{tab:sens_h2h}
\end{table}

\begin{table}[t]
\centering
\begin{tabular}{c|ccccc|}
\cline{2-6}
                                                             & \multicolumn{5}{c|}{\cellcolor[HTML]{EC8F9C}Rating Period}                                                                                                                                                                              \\ \hline
\rowcolor[HTML]{E9E5DC} 
\multicolumn{1}{|c|}{\cellcolor[HTML]{EC8F9C}Sport}          & \multicolumn{1}{c|}{\cellcolor[HTML]{E9E5DC}Annual} & \multicolumn{1}{c|}{\cellcolor[HTML]{E9E5DC}Biannual} & \multicolumn{1}{c|}{\cellcolor[HTML]{E9E5DC}Quarterly} & \multicolumn{1}{c|}{\cellcolor[HTML]{E9E5DC}Bimonthly} & Monthly \\ \hline
\multicolumn{1}{|c|}{\cellcolor[HTML]{E9E5DC}Biathlon}       & \multicolumn{1}{c|}{1.43}                           & \multicolumn{1}{c|}{1.44}                             & \multicolumn{1}{c|}{--}                                & \multicolumn{1}{c|}{--}                                & --      \\ \hline
\multicolumn{1}{|c|}{\cellcolor[HTML]{E9E5DC}Biathlon Relay} & \multicolumn{1}{c|}{0.82}                           & \multicolumn{1}{c|}{0.84}                             & \multicolumn{1}{c|}{--}                                & \multicolumn{1}{c|}{--}                                & --      \\ \hline
\multicolumn{1}{|c|}{\cellcolor[HTML]{E9E5DC}Diving}         & \multicolumn{1}{c|}{1.57}                           & \multicolumn{1}{c|}{1.68}                             & \multicolumn{1}{c|}{--}                                & \multicolumn{1}{c|}{--}                                & --      \\ \hline
\multicolumn{1}{|c|}{\cellcolor[HTML]{E9E5DC}Fencing}        & \multicolumn{1}{c|}{1.05}                           & \multicolumn{1}{c|}{1.07}                             & \multicolumn{1}{c|}{1.07}                              & \multicolumn{1}{c|}{--}                                & --      \\ \hline
\multicolumn{1}{|c|}{\cellcolor[HTML]{E9E5DC}Rugby}          & \multicolumn{1}{c|}{0.97}                           & \multicolumn{1}{c|}{0.97}                             & \multicolumn{1}{c|}{1.06}                              & \multicolumn{1}{c|}{1.06}                              & 1.06    \\ \hline
\end{tabular}
\caption{Average value of diagonal entry of $\sqrt{V_t}$, averaged across athletes and rating periods, using different rating periods.}
\label{tab:sensV}
\end{table}

Overall, while the choice of a rating period theoretically reflects an important bias-variance tradeoff, we found that any reasonable choice led to similar results for the datasets we considered.
Nonetheless, we recommend using the shortest possible rating period if the research goal is to maximize predictive accuracy, and to use slightly longer rating periods if the research goal is to conduct principled inference on the $\btheta_t$ parameters.

\section{Simulation study}
\label{AppB}

In this paper, we make a number of approximations, as described in Appendix \ref{AppA_imp}.
To confirm that these approximations do not significantly hurt performance, we validate our method using a simulation study.
In particular, we evaluate how well the LM-T model can recover the true data-generating parameters $\sigma^2, w$, and $\blambda$.

We simulate data according to the data-generating process implied by the model:
\begin{enumerate}
	\item \textbf{Set data size and data-generating parameters}: pick values for $p$, $T$, $\{g_t\}$, and $\{n_{tg}\}$; also, pick true values for $v_0$, $\sigma^2$, $w$, and $\blambda$ for a true transformation $\tau_\blambda(\cdot)$.
	\item \textbf{Generate athlete latent abilities}: draw $\btheta_1 \sim N(0, \sigma^2 v_0)$ and $\btheta_t \sim N(\btheta_{t-1}, \sigma^2 w)$ for $t = 1, \dots, T$.
	\item \textbf{Generate untransformed scores}: for each game $g$ in rating period $t$, randomly sample $n_{tg}$ athletes and generate untransformed scores as: $\bpsi_{tg} \sim N(\btheta_{tg} - \bar{\btheta}_g, \sigma^2)$.
	\item \textbf{Generate observed scores}: compute $\by_{tg} = t^{-1}_\blambda(\bpsi_{tg})$.
\end{enumerate}
Note that while the simulated untransformed scores are not explicitly game-centered, their mean is zero within each game.

To evaluate parameter recovery and compare model predictions, we simulate 50 datasets for each of a variety of data-generating parameters.
We fix $p=100$ and $T=20$, and consider $n_{tg} = $ 2, 10, and 50 to correspond to head-to-head games and small/large multi-competitor games.
For each setting, we set $g_t$ such that $n_{tg} * g_t$ equals a range of values from 20 to 250 to simulate low to high data environments.
Finally, we fix $v_0 = 10$, $\sigma^2 = 100$, and $w = 0.5$, and we consider data-generating Yeo-Johnson transformations (\cite{Yeo2000}) with $\lambda = 0.7, 1, 1.3$ to see how well the DLM with transformations can recover transformations that shrink, preserve, and stretch the data.
After simulating the multi-competitor data, we run the LM-T model.
We record its estimate of the hyperparameters $(w, \blambda)$ and the posterior mean of $\sigma^2$.

\subsection{Simulation results: parameter recovery}

In this section, we verify that the LM-T model can accurately recover the data-generating parameters in multi-competitor games.

\begin{figure}[ht]
    \centering
    \includegraphics[width=5.5in]{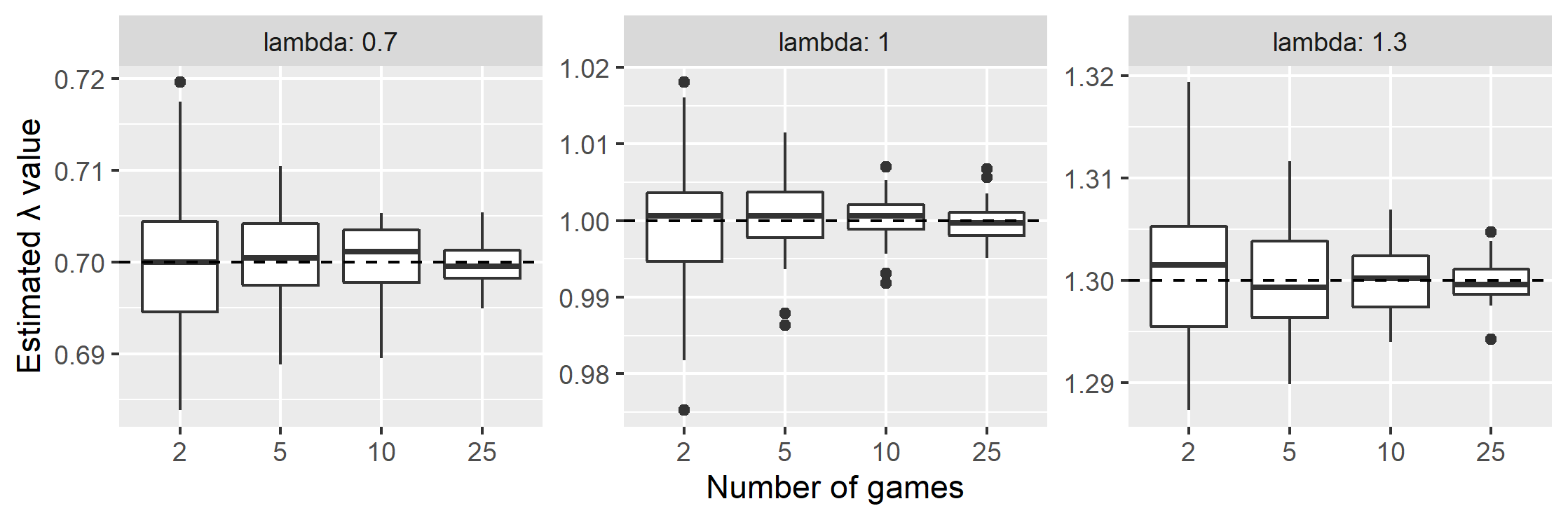}
    \caption{LM-T model recovery of $\lambda$. Dotted line represents the true data-generating $\lambda$ parameter.}
    \label{fig:recover_lambda}
\end{figure}

First, we verify that the LM-T model can accurately recover the $\lambda$ parameter for the data-generating Yeo-Johnson transformation.
Figure \ref{fig:recover_lambda} plots the estimated value of $\lambda$ against the number of multi-competitor games per rating period, aggregated across 50 simulations in each setting.
The dotted lines represent the true data-generating value of $\lambda$.
We see that $\lambda$ recovery is generally extremely accurate; even with only two ten-player games per rating period, the estimated value of $\lambda$ is rarely more than 0.01 away from the true value of $\lambda$.

\begin{figure}[ht]
    \centering
    \includegraphics[width=5.5in]{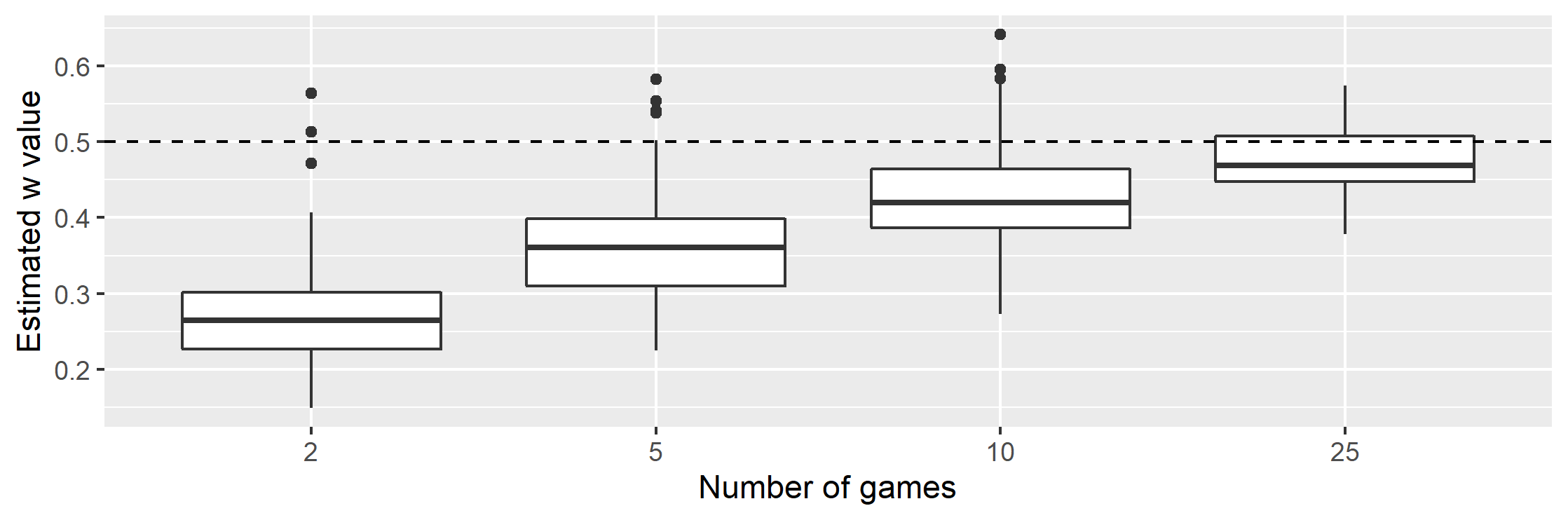}
    \caption{LM-T model recovery of $w$. Dotted line represents the true data-generating $w$ parameter.}
    \label{fig:recover_w}
\end{figure}

Figure \ref{fig:recover_w} plots the value of $w$ estimated by the LM-T model against the number of games per rating period, across 150 datasets.\footnote{We aggregate across the true $\lambda$ values in the simulations, since results for $w$ are qualitatively identical regardless of $\lambda$.}
We see that $w$ recovery is less accurate than $\lambda$ recovery.
The LM-T model generally underestimates $w$ when there are few games, and only consistently recovers the true data-generating $w$ with 25 ten-player games per rating period.
This behavior is appropriate; when there are few games per rating period to accurately estimate player strengths, it makes sense to be more conservative when updating the estimates between rating periods, i.e., to have a lower $w$.

\begin{figure}[ht]
    \centering
    \includegraphics[width=5.5in]{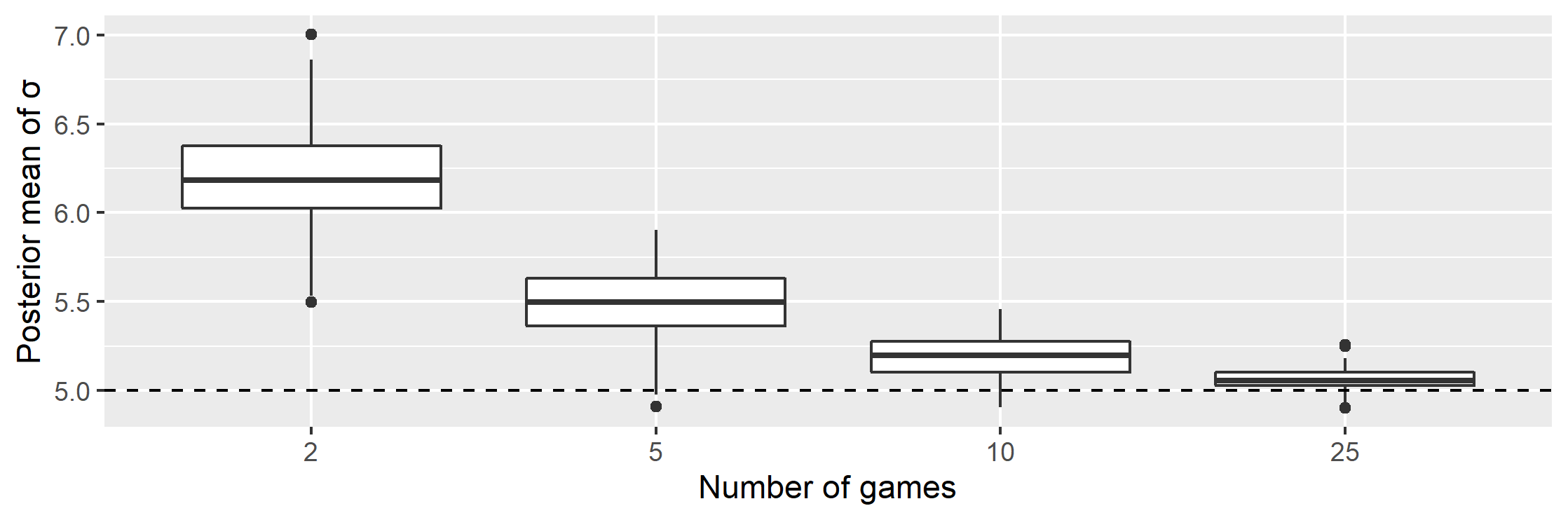}
    \caption{LM-T model recovery of $\sigma$. Dotted line represents the true data-generating $\sigma$ parameter.}
    \label{fig:recover_sigma}
\end{figure}

Similarly, Figure \ref{fig:recover_sigma} plots the value of $\sigma$ estimated by the LM-T model against the number of games per rating period, again across 150 datasets.
We see that $\sigma$ recovery is also less accurate than $\lambda$ recovery for small data sizes, but that the behavior of the $\sigma$ estimates is still appropriate.
When there are few observations per rating period, we cannot be too certain of our estimated latent ability parameters within each rating period; this inflates our $\sigma$ estimate (which in turn deflates our $w$ estimate).
Again, it makes sense for the model to be somewhat conservative when the data do not provide much information.

\end{document}